\documentclass[prd,aps,eqsecnum,floatfix,nofootinbib,preprint,tightenlines,showpacs]{revtex4}

\usepackage{amsmath}
\usepackage{latexsym}
\usepackage{graphicx}

\usepackage[pdftex]{color}
\usepackage[colorlinks=true,linkcolor=blue,filecolor=blue,urlcolor=blue,citecolor=blue,pdftex=true,plainpages=false]{hyperref}

\def\bibi{\bibitem}

\let\inodot=\i

\def\a{\alpha}
\def\b{\beta}
\def\c{\chi}
\def\d{\delta}
\def\f{\phi}                    
\def\g{\gamma}
\def\h{\eta}
\def\i{\iota}

\def\k{\kappa}
\def\l{\lambda}
\def\m{\mu}

\def\p{\pi}                     

\def\F{\Phi}

\def\P{\Pi}

\def\cc{{\cal C}}
\def\cd{{\cal D}}

\def\ch{{\cal H}}   

\def\cl{{\cal L}}

\def\cp{{\cal P}}

\def\ct{{\cal T}}
\def\cu{{\cal U}}

\def\cbo{{\,\raise-.15ex\Sc [\,}}                       

\def\ddt#1{{\buildrel {\hbox{\LARGE .\kern-2pt.}} \over {#1}}}

\def\ie{\mbox{\it i.e.}}
\def\eg{\mbox{\it e.g.}}
\def\etc{\mbox{\it etc.}}

\def\tr{{\rm tr}\,}
\def\Tr{{\rm Tr}\,}

\def\bibi{\bibitem}    

\long\def\symbolfootnote[#1]#2{\begingroup%
\def\thefootnote{\fnsymbol{footnote}}\footnote[#1]{#2}\endgroup}

\long \def \blockcomment #1\endcomment{}

\def\seef{{\it cf.}}

\def\rket#1{| #1\rangle_R}
\def\ChPT{ChPT}
\def\PQChPT{PQChPT}
\newcommand\str{{\rm str\,}}

\def\seef{{\it cf.}}

\def\tb{\tilde{b}}
\def\vx{{\vec x}}
\def\vy{{\vec y}}
\def\vk{{\vec k}}
\def\vl{{\vec\ell}}
\def\vp{{\vec p}}
\def\vq{{\vec q}}
\def\vj{{\vec j}}
\def\vpi{{\vec\p}}
\def\hT{\hat{T}}
\def\hF{\hat{\F}}
\def\hP{\hat{\P}}

\def\bc{\overline{\c}}
\def\hA{\hat{A}}
\def\hB{\hat{B}}
\def\PT{\cp\ct}
\def\hch{{\hat{\ch}}}

\begin{document}

\title{\large\bf On the foundations of partially quenched\\ chiral perturbation theory\\ }

\author{Claude Bernard}
\affiliation{Department of Physics, Washington University, St. Louis, MO 63130, USA}

\author{Maarten Golterman%
\footnote{Permanent address: Department of Physics and Astronomy,
San Francisco State University, San Francisco, CA 94132, USA}}
\affiliation{Institut de F\'\inodot sica d'Altes Energies, 
Universitat Aut\`onoma de Barcelona, E-08193 Bellaterra, Barcelona, Spain}

\date{\today\\ \vspace{0.5cm}}

\begin{abstract}
It has been widely assumed that
partially quenched chiral perturbation theory is the correct low-energy effective theory for partially quenched QCD.   Here we present arguments 
supporting this assumption.   First, we show that, for
partially quenched QCD with staggered quarks, a transfer matrix can
be constructed.  This transfer matrix is not Hermitian, but it is bounded,
and it can be used to construct correlation functions in the usual way.
Combining these observations with an extension of the Vafa--Witten
theorem to the partially quenched theory allows us to argue that the partially quenched
theory satisfies the cluster property.   
By extending Leutwyler's analysis of the
unquenched case to the partially quenched theory, we then conclude that the
existence and properties of the transfer matrix as well as clustering are sufficient
for partially quenched chiral perturbation theory to be the correct low-energy
theory for partially quenched QCD.
\end{abstract}
\maketitle

\newpage
\section{\label{introduction} Introduction}
Partially quenched chiral perturbation theory (\PQChPT) has been extensively
used in the analysis of numerical computations of hadronic quantities in
lattice QCD.  
 In such computations, one has the freedom to vary valence
quark masses (masses of quark operators appearing explicitly in correlation
functions) and sea quark masses (masses of quarks appearing in the fermion
determinant of the theory) independently.   This generalized version of
QCD, which is commonly referred to as partially quenched QCD (PQQCD),
contains full QCD as the special case in which valence and sea
quark masses are set equal to each other (for each flavor) \cite{BGPQ}.
\PQChPT\ is, correspondingly, the generalization of chiral perturbation theory
to the partially quenched setting.\footnote{For a review, see Ref.~\cite{MGLH}.}   

The ability to vary valence and sea quark masses independently is useful
for a variety of reasons.   First, the computation of
quark propagators needed for the contractions making up a correlation
function is significantly less expensive in most applications than the generation of gauge field
configurations, which depend on the sea quark mass.   With fixed computational resources, 
it can thus be an advantage to generate data
for a number of valence quark masses on an ensemble of gauge
configurations with a given sea quark mass.

Second, PQQCD contains full QCD with the same set of sea quarks.
It follows that the low-energy constants (LECs) of the effective theory for the
partially quenched theory are those of the real world, because, by 
definition, the LECs do not depend on the quark masses \cite{ShSh2000}.\footnote{They
do depend on the number of sea-quark flavors.}   It turns out that in a number
of cases, it is easier to determine these LECs by varying the valence and
sea quark masses independently; having more parameters to vary provides
more ``handles'' on the theory.

Third, on the lattice, PQQCD can be generalized to QCD with a ``mixed
action,'' in which not only valence and sea quark masses are 
independently chosen, but also the discretization of the Dirac operator is
different for valence and sea quarks \cite{BRS2003}.   The continuum limit
of such a theory is a partially quenched theory; a fine tuning
is generally required to make valence and sea quark masses equal in the
continuum limit.

Fourth, lattice theories with staggered fermions that use the fourth-root procedure
\cite{Marinari:1981qf}
to eliminate unwanted ``taste'' degrees of freedom are closely related to
PQQCD because there is a mismatch between the (rooted) sea quarks and
the (unrooted) valence quarks \cite{Bernard:2006vv,SharpePoS06,Bernard:2007eh}.  
PQChPT plays an important role in the development of the effective chiral
theory for rooted staggered fermions \cite{Bernard06,BGS08}, which in turn 
provides some evidence for the validity of the rooting procedure. For more discussion
of these issues see Refs.~\cite{SharpePoS06,KronfeldPoS07,GoltermanPoS08,MILC-RMP}
and references therein.

However, it is less clear than in the case of full QCD that the 
partially quenched chiral
theory is indeed the proper low-energy effective field theory.   This is because
PQQCD violates some of the properties of a healthy
quantum field theory:  The path integral definition of PQQCD
includes an integral over ghost quarks, which have the same quantum
numbers and masses as the valence quarks, but which have bosonic,
rather than fermionic, statistics.   The partially quenched theory thus violates
the spin-statistics theorem.
The reason for the presence of ghost quarks is that 
their determinant cancels the valence quark determinant
\cite{Morel};
it is this very cancellation that makes the gauge configurations independent
of the valence mass.

In Ref.~\cite{SW1979}, Weinberg conjectured
that the validity of chiral perturbation theory (\ChPT) as a 
low-energy effective
theory for the Goldstone sector of QCD follows from the
basic properties of a healthy quantum field theory, which include analyticity, 
unitarity, cluster
decomposition, and symmetry considerations.  It was assumed
that the $S$-matrix calculated with
the most general local Lagrangian consistent with a certain symmetry group
is the most general possible $S$-matrix consistent with
these basic properties.  This was then used as a starting point for the systematic
development of \ChPT\ to obtain $S$-matrix elements as an expansion in terms of the pion momenta and masses, following a well-defined power-counting scheme.
The implicit reliance of this argument on unitarity, though, appears to be an obstruction to extending this line of reasoning to the partially quenched case, 
which is certainly not unitary.

A justification for \ChPT\ as the low-energy effective theory for QCD based
on a somewhat different set of arguments
was presented by Leutwyler \cite{HLfound}.   In this justification, the most important
ingredients, in addition to symmetries, are 
locality and the cluster property of the underlying theory (full QCD),
while unitarity is not used.
Locality and clustering
guarantee the existence of vertices in the effective theory that
are independent of the correlation functions in which they appear, and,
consequently, the existence of a loop expansion.  We have
found this approach more useful for the case of PQQCD than that of Ref.~\cite{SW1979}.  By construction, PQQCD is local.   The question then becomes whether the
theory also satisfies the cluster property.   The main goal
of the present article is to collect the theoretical evidence that this is indeed
the case.   Our argument will be based on three ingredients:  the
existence and properties of a transfer matrix for PQQCD, the observation that
chiral symmetry breaking is expected to take place in PQQCD because it
takes place in the full theory contained within PQQCD \cite{ShSh},
and an extension of
the Vafa--Witten theorem \cite{VW} about the absence of spontaneous symmetry 
breaking of vector-like global symmetries (this part of the argument 
extends the argument already given in Ref.~\cite{ShSh}).   The combination of these
ingredients allows us to argue that the partially
quenched theory satisfies the cluster property, under a set of mild additional assumptions similar to those used in Ref.~\cite{HLfound}.   With this result in hand,
the justification for \PQChPT\ as the low-energy effective theory for PQQCD
then follows, much like it does for the case of full QCD.   For technical reasons,
we limit ourselves to lattice QCD with staggered quarks, but we believe that 
the extension to other discretizations of QCD is relatively straightforward.%
\footnote{For instance, see Ref.~\cite{GSS} for Wilson ghost 
quarks.}
An earlier account of part of this work appeared in Ref.~\cite{BGlat10}.

As is well known, the partially quenched theory has more
severe mass singularities than normal QCD. These arise in particular when
valence and ghost masses vanish with sea masses held fixed and nonzero 
(``partially
quenched chiral logarithms'') and are caused by  
double poles in flavor-neutral
propagators \cite{BGPQ,Sharpe:1997by}.  The double poles are 
properties of PQQCD itself, not just of the chiral effective theory, as
shown (with some assumptions) in Ref.~\cite{ShSh}. Here, we will avoid
these mass singularities by always working with all quark and ghost masses
strictly positive.  This has the further advantage that there are no
massless particles (either from chiral symmetry breaking,
or from  breaking of other symmetries, which we show cannot occur).  
Thus ``clustering'' in this article means, ``exponential clustering,'' with
correlation functions of widely separated Euclidean points falling
exponentially with distance. We emphasize that we do not need to take the limit
of vanishing masses in order to show that pseudo-Goldstone bosons (pions) exist in the
partially quenched theory; as mentioned above, this is a consequence of their existence
in unquenched QCD and the (extended) Vafa--Witten theorem.

In constructing the PQQCD transfer matrix, we begin with a theory of
ghost (bosonic) quarks only, coupled to background gluons.  This is clearly the nontrivial part
of the problem, since transfer matrices for ordinary quarks and gluons are standard.  Our 
approach makes it unnatural to demand that ghost-quark and 
valence-quark masses (call them $m_q$ and $m_v$, respectively)
are equal from the outset, as they are in the usual numerical application
of PQQCD.  Thus we are led to a more general setting, where all three types of quark masses
(valance quark, sea quark, and ghost quark) may be different.  This general setting has some  
interesting features, most of which seem to be of purely academic interest, since
the limit of equal ghost-quark and valence-quark masses is the useful one.  However, it
does provide one important insight:  The double poles, which are considered
to be characteristic of PQQCD
and PQChPT, arise from the near-cancelation of single poles in the limit $m_g\to m_v$.
It is crucial here that the poles associated with ghosts have residues with opposite
signs from those of the valence quarks. These unusual signs are associated 
with the bosonic nature of the ghost quarks, which also causes the ghost-quark Hamiltonian to be non-Hermitian.   The ensuing violations of unitarity (see
Sec.~\ref{two-point}) thus appear to be a more fundamental feature of PQQCD 
than the existence of double poles {\it per se}, which appear only in
the special case $m_g=m_v$.

Instead of using bosonic quarks to cancel 
the unwanted valence-quark determinant,
Damgaard and Splittorff \cite{Damgaard:2000gh} proposed a replica approach,
in which each  valence quark is replicated $n_r$ times, and 
one attempts to continue $n_r$ to zero from the positive integers at the end
in order to remove the determinant.  It is interesting to consider
whether further progress in justifying PQChPT can be made using that
approach.  The main obstacle in that direction seems to be the absence
of a proof that the replica approach is indeed equivalent to PQQCD 
nonperturbatively.  If that obstacle were overcome, many other steps would
be straightforward, since the theory has a conventional chiral theory for
each positive integer $n_r$.  However, a similar problem would also remain
on the chiral side of the argument, since one also has no proof that the
replica version of the chiral theory is nonperturbatively
 equivalent to PQChPT 
when $n_r$ is continued to zero.

This article is organized as follows.  In Sec.~\ref{transfer} we construct the transfer matrix for PQQCD with staggered quarks.   We find that the
transfer matrix is not Hermitian, but is nevertheless bounded.   It turns out to be instructive to consider the free theory in some detail, and this is done in Sec.~\ref{free}.  In particular, the free theory clearly demonstrates that
unitarity is violated in PQQCD.   

Then, in Sec.~\ref{effective}, we turn to the effective theory, \PQChPT.
We first give a brief recapitulation of Leutwyler's arguments 
for the unquenched case in Sec.~\ref{Leutwyler}, focusing on the use of 
clustering.  In the partially quenched case,  the 
exponential clustering ``almost''
follows from the existence of a bounded transfer matrix, but the possibility
of massless particle from spontaneous symmetry breaking is a significant loophole.
In Sec.~\ref{VWarg}, we argue that the extension of the
Vafa--Witten theorem about vector-like global symmetries in QCD implies (up to
certain mild assumptions)
that the cluster property does in fact hold in PQQCD.   We discuss
the role of rotational symmetry in Sec.~\ref{O4}, and synthesize all our
observations into an argument for the correctness of \PQChPT\ as the
low-energy effective theory for PQQCD in Sec.~\ref{PQChPTsection}.   
For technical reasons,
we need to assume that the pion masses remain real in PQQCD, despite the fact that
the corresponding transfer-matrix Hamiltonian is not Hermitian.  
This assumption is strongly supported by
numerical evidence. 
In Sec.~\ref{CPTsection}, we use CPT symmetry to argue
that all low energy
constants in PQChPT have the same phases (real, with usual conventions) as
they would in a Hermitian theory. This supports our assumption that the
masses are real. Nevertheless, as we show in Sec.~\ref{non-degenerate}, in the 
nonstandard case where ghost-quark masses are not 
degenerate with valence-quark masses, the effective theory shows that complex masses
may arise in some ranges of quark and ghost masses, as may a new phase transition.
The effective theory also shows how double poles arise from single poles in the
$m_g\to m_v$ limit.
Our conclusions are contained in Sec.~\ref{conclusion}.   There are two
appendices, both concerned with the free theory.   In App.~\ref{completeness}
we show completeness of a basis of (right or left) eigenstates, and App.~\ref{path}
discusses a path integral formulation of the free theory.

\section{\label{transfer} Transfer matrix}
In this section, we construct the transfer matrix for a gauge theory
coupled to fermionic and bosonic or ``ghost'' staggered quarks.   Each of these
quarks can have an arbitrary mass, but we will require all
masses to be positive.\footnote{The phase of a ghost-quark mass cannot
be changed by chiral transformations \cite{GSS,DOTV}.}
   In Sec.~\ref{ghosts} we construct the transfer
matrix for ghost quarks in a background gauge field.   In Sec.~\ref{properties}
we discuss some properties of the ghost transfer matrix; in particular,
we show that it is bounded.  Then, 
in Sec.~\ref{complete} we combine this with the transfer matrix
for a gauge theory with only fermionic quarks to arrive at the
complete transfer matrix, and show that it is invariant under PT and CPT
symmetry.  We will use the ``double time-slice''
construction for both fermionic and ghost quarks \cite{vdDS,Smit}.

\subsection{\label{ghosts} Ghost sector}
The staggered action is
\begin{equation}
\label{stagS}
S=\sum_x\left\{\frac{1}{2}\sum_\m\h_\m(x)\left(\c^\dagger(x)U_\m(x)\c(x+\m)-
\c^\dagger(x+\m)U_\m^\dagger(x)\c(x)\right)
+m\c^\dagger(x)\c(x)\right\}\ ,
\end{equation}
in which 
\begin{equation}
\label{etas}
\h_\m(x)=(-1)^{x_1+\dots x_{\m-1}}\ ,
\end{equation}
and where 
$\c$ and $\c^\dagger$ are the staggered fields, $U_\m(x)$ are the link variables, 
and color indices are suppressed.    We will denote a lattice
gauge field consisting of all link variables $U_\m(x)$ by $\cu$.
If $\c$ and $\c^\dagger$ are Grassmann, they are independent of each other,
and $\c^\dagger$ can then also be denoted as $\bc$, as is often done.   However, here we
are interested in the ghost-quark sector of the theory, for which we take
$\c(x)$ to be a $c$-number.   In order that the path integral over the
ghost fields be convergent,
we need to take $\c^\dagger(x)$ to be the Hermitian
conjugate of $\c(x)$.    
With this choice, the partition function
\begin{equation}
\label{Z}
Z(\cu)=\int\prod_x d\c^\dagger_xd\c_x\;\mbox{exp}\left(-S\right)
\end{equation}
is well-defined, as long as we take $m>0$, which we will assume throughout
this article. 

In order to construct the transfer matrix representation of $Z$, we find it
convenient to introduce real fields $\phi_1$ and $\phi_2$ through
\begin{equation}
\label{real}
\c(x)=\h_4(x)\f_1(x)+i\f_2(x)\;,\qquad\c^\dagger(x)=\h_4(x)\f_1(x)-i\f_2(x)\ .
\end{equation}
Choosing temporal gauge, $U_4(x)=1$, and
splitting $x\to (\vx,\,t= x_4)$, we rewrite $S$ as\footnote{The fields $\phi_1$ and $\phi_2$
are always transposed when they appear as the first factor in a bilinear
in these fields.  We also use that $\eta_4(\vx+\vj)=-\eta_4(\vx)$.}
\begin{eqnarray}
\label{Sinphi}
S&=&\sum_x\Biggl\{i\Bigl(\f_1(\vx,t)\f_2(\vx,t+1)-\f_2(\vx,t)\f_1(\vx,t+1)\Bigr)\\
&&\hspace{-1cm}+i\sum_j\h'_j(\vx)\left(\f_1(\vx,t)\;\mbox{Re}\;U_j(\vx,t)\f_2(\vx+\vj,t)
+ \f_2(\vx,t)\;\mbox{Re}\;U_j(\vx,t)\f_1(\vx+\vj,t)\right)\nonumber\\
&&\hspace{-1cm}+i\sum_j\h_j(\vx)\left(-\f_1(\vx,t)\;\mbox{Im}\;U_j(\vx,t)\f_1(\vx+\vj,t)
+\f_2(\vx,t)\;\mbox{Im}\;U_j(\vx,t)\f_2(\vx+\vj,t)\right)\nonumber\\
&&\hspace{0.7cm}+m\left(\f_1(x)^2+\f_2(x)^2\right)\Biggr\}\ ,\nonumber
\end{eqnarray}
in which
\begin{equation}
\label{etap}
\h'_j(\vx)=\h_j(\vx)\h_4(\vx)\ ,
\end{equation}
and
\begin{eqnarray}
\label{ReIm}
\mbox{Re}\;U_k(x)&=&\frac{1}{2}(U_k(x)+U^*_k(x))\ ,\\
\mbox{Im}\;U_k(x)&=&\frac{1}{2i}(U_k(x)-U^*_k(x))\ .\nonumber
\end{eqnarray}
Next, we divide the lattice into even and odd time slices,
and rename the fields $\Phi$, respectively $\Pi$, defining
\begin{eqnarray}
\label{timeslices}
&\hspace{-0.7cm} t=2k:&\f_1(\vx,t)=\F_{1,k}(\vx)\ ,\ \ \ \ \ \f_2(\vx,t)=-\F_{2,k}(\vx)\ ,\\
&t=2k+1:&\f_1(\vx,t)=\P_{2,k}(\vx)\ ,\ \ \ \ \ \f_2(\vx,t)=\P_{1,k}(\vx)\ .\nonumber
\end{eqnarray}
Accordingly, the action can be rewritten as
\begin{eqnarray}
\label{Sts}
S&=&\sum_k\Biggl\{\sum_\vx i\left(\F_{1,k}(\vx)\P_{1,k}(\vx)+\F_{2,k}(\vx)\P_{2,k}(\vx)\right)\\
&&
\qquad+\ch_-[\Phi_{1,k},\Phi_{2,k};\cu(2k)]+\ch_0[\Phi_1,\Phi_2;m]\Biggr\}\nonumber\\
&&+\sum_k\Biggl\{\sum_\vx-i\left(\P_{1,k}(\vx)\F_{1,k+1}(\vx)+\P_{2,k}(\vx)\F_{2,k+1}(\vx)\right)\nonumber\\
&&\qquad+\ch_+[\Pi_{1,k},\Pi_{2,k};\cu(2k+1)]+\ch_0[\Pi_1,\Pi_2;m]\Biggr\}\ ,
\nonumber
\end{eqnarray}
in which $\cu(t)$ denotes the gauge field at a time slice $t$, and
\begin{eqnarray}
\ch_\pm[\Psi_1, \Psi_2;\; \cu(t)]&=&\sum_\vx\Biggl\{\pm\sum_j i\h'_j(\vx)\left(\Psi_1(\vx)\;\mbox{Re}\;
U_j(\vx,t)\Psi_2(\vx+\vj)+(1\leftrightarrow 2)\right)\nonumber\\
&&\hspace{+0.75cm}-\sum_j i\h_j(\vx)\left(\Psi_1(\vx)\;\mbox{Im}\;
U_j(\vx,t)\Psi_1(\vx+\vj)-(1\to 2)\right)
\Biggr\}\ ,\nonumber\\
\ch_0[\Psi_1,\Psi_2;m]&=&m\sum_\vx\left(\Psi_{1,k}(\vx)^2+\Psi_{2,k}(\vx)^2\right)\ .
\label{chdef}
\end{eqnarray}
After shifting $k$ to $k+1$ in the $\ch_-$ term, we define a kernel
\begin{eqnarray}
\label{kernel}
\hspace{-0.3cm}T(\F_{1,k+1}\F_{2,k+1};\F_{1,k}\F_{2,k})&\!=\!&\int \prod_\vy d\P_{1,k}(\vy)\int \prod_\vy d\P_{2,k}(\vy)
\,\times\\
&&\hspace{-5.2cm}
\mbox{exp}\Biggl[-\Biggl\{\sum_\vx i\left(\F_{1,k}(\vx)\P_{1,k}(\vx)\!+\!\F_{2,k}(\vx)\P_{2,k}(\vx)
\!-\!\P_{1,k}(\vx)\F_{1,k+1}(\vx)\!-\!\P_{2,k}(\vx)\F_{2,k+1}(\vx)\right)\nonumber\\
&&\hspace{-3.8cm}+ \ch_-[\F_{1,k+1},\F_{2,k+1};\;\cu(2(k+1)]+\ch_+[\P_{2,k},\P_{1,k};\;\cu(2k+1)]\nonumber\\
&&\hspace{1cm}+\ch_0[\Phi_1,\Phi_2;m]+\ch_0[\Pi_1,\Pi_2;m]\Biggr\}\Biggr]\ .\nonumber
\end{eqnarray}
The claim is then that
\begin{equation}
\label{connection}
T(\F_{1,k+1}\F_{2,k+1};\F_{1,k}\F_{2,k})=\langle\F_{1,k+1}\F_{2,k+1}|\hT_{G,k}(\cu)|\F_{1,k}\F_{2,k}\rangle\ ,
\end{equation}
with
\begin{equation}
\label{tmatrix}
\hT_{G,k}(\cu)=e^{-\ch_-[\hF_1,\hF_2;\;\cu(2(k+1)]-
\ch_0[\hF_1,\hF_2;m]}\,
e^{-\ch_+[\hP_2,\hP_1;\;\cu(2k+1)]-\ch_0[\hP_1,\hP_2;m]}\ ,
\end{equation}
in which the Hermitian operators $\hF_a(\vx)$ and $\hP_a(\vx)$ obey the commutation
rules
\begin{equation}
\label{commrules}
[\hF_a(\vx),\hP_b(\vy)]=i\d_{\vx,\vy}\;\d_{ab}\ .
\end{equation}
This is proven by inserting a complete set of states into the right-hand side of Eq.~(\ref{connection}):
\begin{eqnarray}
\label{proof}
\langle\F_{1,k+1}\F_{2,k+1}|\hT_{G,k}(\cu)|\F_{1,k}\F_{2,k}\rangle&=&\\
&&\hspace{-6cm}
\int d\P_{1,k}\int d\P_{2,k}\;\langle\F_{1,k+1}\F_{2,k+1}|
e^{-\ch_-[\hF_1,\hF_2;\;\cu(2(k+1))]}\,e^{-\ch_+[\hP_2,\hP_1;\;\cu(2k+1)]}|\P_{1,k}\P_{2,k}\rangle\nonumber\\
&&\times\ 
\langle\P_{1,k}\P_{2,k}|\F_{1,k}\F_{2,k}\rangle\nonumber\\
&&\hspace{-6cm}=
\int d\P_{1,k}\int d\P_{2,k}\;
e^{-\ch_-[\F_{1,k+1},\F_{2,k+1};\;\cu(2(k+1))]}\,e^{-\ch_+[\P_{2,k},\P_{1,k};\;\cu(2k+1)]}\nonumber\\
&&\times\ 
e^{i\left(\F_{1,k+1}\P_{1,k}+\F_{2,k+1}\P_{2,k}-\F_{1,k}\P_{1,k}-\F_{2,k}\P_{2,k}\right)}\ ,
\nonumber
\end{eqnarray}
where we omitted the explicit arguments $\vx$ from the fields.  Restoring these, the last
line of Eq.~(\ref{proof}) coincides precisely with Eq.~(\ref{kernel}).   The ghost partition function
in a gauge-field background $\cu$ is now given by
\begin{equation}
\label{ghostZ}
Z_G(\cu)=\Tr\left(\prod_{k=1}^{T/2}\hT_{G,k}(\cu)\right)\ ,
\end{equation}
if $T$ is the time extent of the lattice.

\subsection{\label{properties} Properties of the ghost transfer matrix}
The transfer matrix~(\ref{tmatrix}) is not Hermitian, but it is bounded.
The proof is as follows.   The operator $\ch_-[\hF_1,\hF_2;\cu(2(k+1))]
+\ch_0[\hF_1,\hF_2;m]$ consists of a positive semi-definite part
(if $m>0$), and an anti-Hermitian part containing the gauge field.
Moreover, it commutes with its Hermitian conjugate, because all
$\hF_a(\vx)$ commute among themselves.  Therefore, the
operator
\begin{equation}
\label{T1}
\hT_1(\cu)=e^{-\ch_-[\hF_1,\hF_2;\;\cu(2(k+1)]-\ch_0[\hF_1,\hF_2;m]}
\end{equation}
is normal and bounded,\footnote{We use the Euclidean
norm, which, for a matrix $A$, is defined as the positive square root of the
largest eigenvalue of $A^\dagger A$.}
\begin{equation}
\label{T1bound}
\parallel\hT_1(\cu)\parallel\,\le 1\ .
\end{equation}
A similar argument applies to
\begin{equation}
\label{T2}
\hT_2(\cu)=e^{-\ch_+[\hP_2,\hP_1;\;\cu(2k+1)]-\ch_0[\hP_1,\hP_2;m]}\ ,
\end{equation}
and thus it follows that $\hT_G$ itself is bounded,
\begin{equation}
\label{Tbound}
\parallel\hT_G(\cu)\parallel\,\le\,\parallel\hT_1(\cu)\parallel\parallel\hT_2(\cu)\parallel\,\le 1\ .
\end{equation}
This establishes that all eigenvalues of $\hT_G$ have an absolute value less 
than or equal to one.  If, moreover, the eigenvalue $\l_0$ with maximal
absolute value is unique, correlation functions in this theory decay exponentially
with distance.

The transfer matrix~(\ref{tmatrix}) may be assumed to
have a complete set of right and left eigenstates. This is equivalent
to saying that,   
if we block-diagonalize $\hT_G$ (put it in Jordan normal form), there are no blocks of the form
\begin{equation}
\label{Jordan}
\begin{pmatrix}\l&\k\\ 0&\l\end{pmatrix}\ ,
\end{equation}
with $\k\ne 0$ (or generalizations of this form with higher degeneracies).
We will see in Sec.~\ref{free} (and App.~\ref{completeness}) that this situation does not occur in the free
theory.  The condition that it happens on some nontrivial gauge field  in the
interacting case then restricts such fields to a subspace of co-dimension one 
(or more) in the full space of gauge-field
configurations.    Therefore, for ``most'' (meaning all gauge fields except
a set of zero measure in the space of all gauge fields), the transfer matrix
$\hT_{G,k}(\cu)$ can be completely diagonalized, and  complete sets of
right and left eigenstates exist.

There is a caveat, however.   In the next section, we will incorporate the
ghost transfer matrix we have constructed thus far into a  transfer matrix for the
entire theory, including quantized quarks and gauge fields as well as ghosts.   In order
to conclude that the total transfer matrix does not have any nontrivial
Jordan blocks of the form~(\ref{Jordan}) (or generalizations thereof,
\seef\ App.~\ref{completeness}), we would have to construct the full
Hilbert space for the entire matrix, something we do not know
how to do.\footnote{For the free
theory it is obvious that it can be completely diagonalized, because in that case the Hilbert space is the
direct product of the free Hilbert spaces for the quark and ghost
sectors.}    One might, however,
consider a hybrid construction of the entire theory, in which fermions
and ghosts are treated in a canonical formalism, and gauge fields
are taken into account through the path integral.  What this means is
that correlation functions with quarks or ghosts on the external lines
are first constructed in the transfer-matrix formalism, in an arbitrary
fixed gauge-field background.   The full QCD correlation functions are
then obtained by integrating these correlation functions over the gauge
fields.   We first exclude from this integral the measure-zero set of gauge
fields for which the Jordan normal form of the transfer matrix in a fixed
gauge-field background may be nondiagonal. Since we then have a complete
set of eigenstates in each background field, we think it
reasonable to assume that the entire transfer matrix has a complete
set of eigenstates.
\subsection{\label{complete} Full transfer matrix}
It is straightforward to combine the transfer matrix for ghost quarks
constructed in Sec.~\ref{ghosts} with the transfer matrix for lattice QCD with
staggered fermions.   First, from Refs.~\cite{vdDS,Smit}, the fermionic transfer
matrix for a staggered quark in  a fixed background
gauge field $\cu$ can be written in the form
\begin{equation}
\label{fermt}
\hT_{F,k}(\cu)=e^{\hA^\dagger[\cu(2(k+1))]}\,e^{\hB[m]}\,
e^{\hA[\cu(2k+1)]}\ ,
\end{equation}
with $\hB$ Hermitian.
This translates the system from double time slice $k$
(\seef\ Eq.~(\ref{timeslices})), with gauge fields
$\cu(2k)$ and $\cu(2k+1)$, to the next double time slice
$k+1$, with gauge fields $\cu(2(k+1))$ and $\cu(2(k+1)+1)$.  In more
detail, the factor $e^{\hA}$ takes care of the hop within the double slice,
connecting slice $2k$ to slice $2k+1$ (and it  contains spatial terms
on slice $2k+1$), whereas the factor $e^{\hA^\dagger}$
hops from slice $2k+1$ to slice $2(k+1)$ in the next double
slice (and it  contains spatial terms
on the slice $2(k+1)$). 

The ghost transfer matrix of Eq.~(\ref{tmatrix}) can be written similarly as:
\begin{equation}
\label{rewritetm}
\hT_{G,k}(\cu)=e^{-\ch_-[\hF_1,\hF_2;\;\cu(2(k+1))]}\,e^{-\hch[m]}\,e^{-\ch_+[\hP_1,\hP_2;\;\cu(2k+1)]}\ ,
\end{equation}
where
\begin{equation}
\label{h}
e^{-\hch[m]}\equiv e^{-\ch_0[\hF_1,\hF_2;m]}\,e^{-\ch_0[\hP_1,\hP_2;m]}\ .
\end{equation}

We can then write the transfer matrix 
for the total theory of 
QCD with staggered quarks and ghost quarks in the form
\begin{equation}
\label{stagt}
\hT_{total}=\hT_U^{1/2}\,e^{\hA^\dagger}\,e^{-\ch_-[\hF_1,\hF_2
{;\,\hat\cu} ]}\,\hT_U\,e^{-\hch[m]}\,e^{\hB}\,
e^{\hA}\,e^{-\ch_+[\hP_1,\hP_2{;\,\hat\cu}]}\,\hT_U^{1/2}\ ,
\end{equation}
with $\hT_U$ the transfer matrix of the pure gauge theory constructed
in Ref.~\cite{gT}.  The gauge field transfer matrix hops between single time slices, and therefore
a factor $\hT_U$ needs to be inserted between 
$e^{\hA}\,e^{-\ch_+[\hP_1,\hP_2{;\,\hat\cu}]}$ and 
$e^{\hA^\dagger}\,e^{-\ch_-[\hF_1,\hF_2{;\,\hat\cu}]}$, with matrix elements 
\begin{equation}
\label{me}
\langle\cu(2(k+1))|\hT_U|\cu(2k+1)\rangle\ .
\end{equation}
Then, when one writes the partition function as a trace over a power of the
transfer matrix, the factors $\hT_U^{1/2}$ combine to hop the gauge field
from slice $2k$ to $2k+1$ (on the right of Eq.~(\ref{stagt})), or from slice
$2(k+1)$ to slice $2(k+1)+1$  (on the left of Eq.~(\ref{stagt})).

Although $\hT_{total}$ is not Hermitian, and therefore is not required to have only
real eigenvalues, there are significant restrictions on the eigenvalues coming from
discrete symmetries.  Most importantly, $\hT_{total}$ is invariant under the
antiunitary symmetry PT, the product of parity and time-reversal symmetries.
Under PT, the ghost fields in Eq.~(\ref{stagt}) transform according according to
\begin{eqnarray}
\PT\; \hat\Phi_1(\vec x)\; (\PT)^\dagger &=& \hat\Phi_2(-\vec x)\ , \nonumber \\
\PT\; \hat\Phi_2(\vec x)\; (\PT)^\dagger &=& \hat\Phi_1(-\vec x)\ , \nonumber \\
\PT\; \hat\Pi_1(\vec x)\; (\PT)^\dagger &=& -\hat\Pi_2(-\vec x)\ , \label{PT} \\
\PT\; \hat\Pi_2(\vec x)\; (\PT)^\dagger &=& -\hat\Pi_1(-\vec x)\ , \nonumber
\end{eqnarray}
where $\PT$ is the antiunitary operator that generates the symmetry.
Note that, because it is antiunitary, PT it leaves the commutation rules,
Eq.~(\ref{commrules}),
unchanged.
In the free case  ($U_j(\vec x,t)=1$), the invariance of the ghost part of
the transfer matrix, $\hT_G$, under PT can be easily
checked using the definitions of $\ch_{\pm}$ and $\ch_0$, Eq.~(\ref{chdef}), and
the relations $\eta_j(\vec x+\vec j) = \eta_j(\vec x)$
and $\eta'_j(\vec x+\vec j) = -\eta'_j(\vec x)$.  In the interacting case,
the ghost transfer matrix $\hT_G(\cu)$ is of course
not invariant under Eq.~(\ref{PT})
for fixed background fields $U_j(\vec x,t)$; we must also let the time-slice
gauge field operators transform.
The transformation rule is standard:
\begin{equation}
\label{PT-U}
\PT\; \hat U_j(\vec x)\;  (\PT)^\dagger = \hat U_{-j}(-\vec x)\equiv \hat U^\dagger_j(-\vec x-\vec j)\ .
\end{equation}
Of course the quark and pure gauge parts of  $\hT_{total}$ are also invariant under PT.

Since $\hT_{total}$ is invariant under PT,
\begin{equation}
\label{PTT}
\PT\; \hT_{total}\; (\PT)^\dagger = \hT_{total}\ , 
\end{equation}
each eigenvalue of $\hT_{total}$ must be either real or one of a complex conjugate
pair:
\begin{equation}
\hT_{total} \vert \Psi \rangle = \lambda\vert \Psi \rangle \quad\Rightarrow \quad
\hT_{total} (\PT \vert \Psi \rangle) = \lambda^*  (\PT \vert \Psi \rangle ) \ .\label{PTeigenvalues}
\end{equation}

Non-Hermitian Hamiltonians with PT symmetry
have been studied extensively by Bender and others
\cite{Bender}.
PT-symmetric theories 
may often be redefined to give an acceptable unitary theory.  Here, however,
we do not want to make any such redefinitions, since the path integral is given, and the
unitarity violations due to a non-Hermitian transfer matrix (or Hamiltonian) are as
expected for a theory with spin-1/2 bosons.

For future reference, we note that $\hT_{total}$ is also invariant
under charge conjugation symmetry, C, and hence under the combined symmetry
CPT, 
with
\begin{eqnarray}
\cc\PT\; \hat\Phi_1(\vec x)\; (\cc\PT)^\dagger &=& \hat\Phi_1(-\vec x)\ , \nonumber \\
\cc\PT\; \hat\Phi_2(\vec x)\; (\cc\PT)^\dagger &=& \hat\Phi_2(-\vec x)\ , \nonumber \\
\cc\PT\; \hat\Pi_1(\vec x)\; (\cc\PT)^\dagger &=& -\hat\Pi_1(-\vec x)\ , \label{CPT-ghost-op} \\
\cc\PT\; \hat\Pi_2(\vec x)\; (\cc\PT)^\dagger &=& -\hat\Pi_2(-\vec x)\ . \nonumber 
\end{eqnarray}
The gauge field transforms under CPT as
\begin{equation}\label{CPT-gauge}
\cc\PT\; \hat U_j(\vec x)\;  (\cc\PT)^\dagger = \hat U^{{\rm T}}_j(-\vec x-\vec j)\ ,
\end{equation}
with T the matrix transpose.   

From Eq.~(\ref{CPT-ghost-op}), we can find the CPT transformation 
rules for
the c-number fields $\chi(x)$ and $\chi^\dagger(x)$, which are, from Eqs.~(\ref{timeslices})
and ~(\ref{real}), the eigenvalues of (linear combinations of) the operators 
$\hat\Phi_1(\vec x)$, $\hat\Phi_2(\vec x)$, $\hat\Pi_1(\vec x)$, and 
$\hat\Pi_2(\vec x)$ on each time slice. 
Note that, since the time-translation operator in
Euclidean space is $\exp(-H x_4)$, we should not send $ x_4\to - x_4$ under 
this symmetry if we want the Euclidean action, as opposed to merely
the Hamiltonian, to be invariant.%
\footnote{There is however a {\it linear}\/
time-inversion symmetry of the action in Euclidean space, which is simply the time-direction
equivalent of the 
spatial inversions.  We focus instead on the antiunitary symmetry
of the Hamiltonian because it will be useful to us in constraining the chiral
theory.}
We then have
\begin{eqnarray}
{\rm CPT:}\quad \chi(\vec x,x_4) &\to& (-1)^{x_4}\;\chi^{\dagger{\rm T}}(-\vec x,x_4)\ , \nonumber \\
{\rm CPT:}\quad \chi^\dagger(\vec x,x_4) &\to& (-1)^{x_4}\;\chi^{\rm T}(-\vec x,x_4) 
\ , \label{CPT-ghost} 
\end{eqnarray}
where the factors of $(-1)^{x_4}$ arise from the minus signs in the last
two equations in Eq.~(\ref{CPT-ghost-op}).  It is straightforward to check that
the action, 
Eq.~(\ref{stagS}), is unchanged by this transformation.

For staggered quarks, the action is 
\begin{equation}
\label{stag-quarkS}
S=\sum_x\left\{\frac{1}{2}\sum_\m\h_\m(x)\left(\bar q(x)U_\m(x)q(x+\m)-
\bar q(x+\m)U_\m^\dagger(x)q(x)\right)
+m\bar q(x)q(x)\right\}\ ,
\end{equation}
where now $q(x)$ and $\bar q(x)$ are independent Grassmann-valued fields.  Starting
from the naive-quark transfer matrix Hamiltonian  \cite{Smit} or the ``reduced-staggered''
transfer matrix in \cite{vdDS}, one can derive
the transformation rules for $q(x)$ and $\bar q(x)$ that correspond to Eq.~(\ref{CPT-ghost}): 
\begin{eqnarray}
{\rm CPT:}\quad q(\vec x,x_4) &\to& (-1)^{x_4}\;\bar q^{\rm T}(-\vec x,x_4)\ , \nonumber \\
{\rm CPT:}\quad \bar q(\vec x,x_4) &\to& -(-1)^{x_4}\;q^{\rm T}(-\vec x,x_4) 
\ . \label{CPT-quark}
\end{eqnarray}
The extra minus sign in the second equation in Eq.~(\ref{CPT-quark}) (as compared to 
Eq.~(\ref{CPT-ghost})) makes
up for the minus sign coming from Fermi statistics when taking the transpose of the
action.

We end this section with a few comments.   First, implicitly, we have only
considered one flavor of quarks and one flavor of ghost quarks.   The
generalization to arbitrary numbers of each is immediate.
In addition, we have generalized beyond PQQCD by 
choosing an arbitrary (positive) mass for each quark or ghost quark.
In PQQCD, each ghost quark mass is equal to the mass of
one of the fermionic quarks, thus turning that quark into a valence quark.
Sea quarks appear as fermionic quarks without ghost partners.
A further issue arises from the use of the 
staggered action (without rooting) for
the ghosts and quarks, which implies that each  flavor comes in four tastes.  
This is actually not a serious
restriction for PQQCD, since extra (unwanted) species of valence quarks and ghosts are harmless:
they have no effect on any processes if we choose not to put them on external lines
(in the standard case where valence quarks and ghosts are degenerate).  Further, if
one wishes to avoid extra tastes in the sea, it would
be completely straightforward to use any alternative
discretization for the sea quarks that has a transfer matrix, such as unimproved Wilson quarks.

\section{\label{free} The free theory}
This section focuses on the ghost transfer matrix, Eq.~(\ref{tmatrix}), 
in the free theory, {\it i.e.}, in the case where the background gauge
field $U_j(\vx,t)=1$. We work in the limit of vanishing temporal lattice spacing. 
We first follow the standard momentum-space  construction for
staggered fermions \cite{vdDS,GS}
to identify the eight degrees of freedom that arise from spatial doubling (the
doubling associated with the time direction is already explicitly taken into account
in our two-time-slice construction of the transfer matrix).%
\footnote{The 16 degrees of freedom are identified as the four tastes of Dirac
fermions, each with four spin degrees of freedom.}
 We diagonalize the
Hamiltonian in spin-taste space, and proceed to determine the eigenstates and eigenvalues
using a generalized Bogoliubov transformation.  Two-point correlators can then
be easily found; they clearly show the expected violations of unitarity 
in the ghost sector of the theory.
The explicit calculations below and in App.~\ref{completeness} demonstrate that
$\hT_G$ has a complete
set of (right or left) eigenstates in the free theory. In other words, they show
that, when $U_j(\vx,t)=1$, blocks of the form of Eq.~(\ref{Jordan}) do not occur in the Jordan
normal form of the transfer matrix.
As an alternative to the Bogoliubov-transformation approach, a path-integral 
construction of the correlators is given in App.~\ref{path}.

\subsection{\label{Hamiltonian} The free Hamiltonian}
Setting $U_j(\vx,t)=1$, writing\footnote{The factor 2 appears because
$\hT_G$ is a double time-slice transfer matrix.}
\begin{equation}
\label{TH}
\hT_{G,k}(\cu=1)\equiv\mbox{exp}\left(-2a_t H\right) \ ,
\end{equation}
and taking the limit $a_t\to 0$,
we find the Hamiltonian $H$ in that limit:\footnote{From
now on we drop hats on operators.}
\begin{eqnarray}
\label{ham}
H&=&\frac{1}{2}\sum_\vx\Biggl\{
m\left(\F_1(\vx)^2+\F_2(\vx)^2+\P_1(\vx)^2+\P_2(\vx)^2\right)+\sum_ji\h'_j(\vx)\,\times\\
&&\left(-\F_1(\vx)\F_2(\vx+\vj)-\F_1(\vx+\vj)\F_2(\vx)
+\P_2(\vx)\P_1(\vx+\vj)+\P_2(\vx+\vj)\P_1(\vx)\right)\Biggr\}\;.\nonumber
\end{eqnarray}
Introducing creation and annihilation operators $a_1^\dagger$, $a_1$, $a_2^\dagger$, $a_2$ through
\begin{eqnarray}
\label{crann}
\F_1(\vx)&=&\int_k\frac{1}{\sqrt{2}}\left(a_1(\vk)+a_1^\dagger(-\vk)\right)e^{i\vk\cdot\vx}\ ,\\
\P_1(\vx)&=&\int_k\frac{-i}{\sqrt{2}}\left(a_1(\vk)-a_1^\dagger(-\vk)\right)e^{i\vk\cdot\vx}\ ,\nonumber\\
\F_2(\vx)&=&\int_k\frac{-i}{\sqrt{2}}\left(a_2(-\vk)-a_2^\dagger(\vk)\right)e^{i\vk\cdot\vx}\ ,\nonumber\\
\P_2(\vx)&=&\int_k\frac{-1}{\sqrt{2}}\left(a_2(-\vk)+a_2^\dagger(\vk)\right)e^{i\vk\cdot\vx}\ ,
\nonumber
\end{eqnarray}
in which
\begin{equation}
\label{momint}
\int_k\equiv\int\frac{d^3k}{(2\p)^{\frac{3}{2}}}\ ,
\end{equation}
this can be re-expressed as
\begin{eqnarray}
\label{Hcrann}
H&=&\frac{1}{2}\sum_\vx\int_k\int_\ell\;e^{i(\vk+\vl)\cdot\vx}\,\times\\
&&\hspace{-0.5cm}\Biggl\{m\left(a_1^\dagger(-\vk)a_1(\vl)+a_1(\vk)a_1^\dagger(-\vl)+
a_2^\dagger(\vk)a_2(-\vl)+a_2(-\vk)a_2^\dagger(\vl)\right)\nonumber\\
&&\hspace{-0.5cm}+\sum_j\;e^{i\ell_j}\;e^{i\vpi_{\h'_j}\cdot\vx}\left(
-a_1(\vk)a_2(-\vl)+a_1^\dagger(-\vk)a_2^\dagger(\vl)
-a_2(-\vk)a_1(\vl)+a_2^\dagger(\vk)a_1^\dagger(-\vl)\right)\Biggr\}\ .\nonumber
\end{eqnarray}
Here the factors $e^{i\vpi_{\h'_j}\cdot\vx}$ are equal to the sign factors $\h'_j(\vx)$
in Eq.~(\ref{ham}), if we choose
\begin{equation}
\label{pis}
\vpi_{\h'_1}=(\p,\p,\p)\ ,\ \ \ \vpi_{\h'_2}=(0,\p,\p)\ , \ \ \ \vpi_{\h'_3}=(0,0,\p)\ .
\end{equation}
The creation and annihilation operators have commutation rules
\begin{equation}
\label{cranncomm}
[a_\a(\vk),a_\b^\dagger(\vl)]=\d(\vk-\vl)\;\d_{\a\b}\ .
\end{equation}

We now split up the (spatial) Brillouin zone as in Refs.~\cite{vdDS,GS} for staggered fermions:\footnote{See also Ref.~\cite{MGLH} for a review.}
\begin{equation}
\label{split}
\vk=\vp+\vpi_A\ ,\ \ \ \ \ \vl=\vq+\vpi_B\ ,
\end{equation}
with
\begin{equation}
\label{AB}
\vpi_A,\vpi_B\in\{(0,0,0),(\p,0,0),\dots,(\p,\p,\p)\}\ ,
\end{equation}
such that $-\p/2<p_j,q_j\le\p/2$.
Operators get relabeled as in
\begin{equation}
\label{relabel}
a_\a(\vk)=a_\a(\vp+\vpi_A)\equiv a_\a^A(\vp)\ ,
\end{equation}
\etc \ Performing the sum over $\vx$ in Eq.~(\ref{Hcrann}) we find the delta functions
\begin{eqnarray}
\label{delta}
\d(\vp+\vq+\vpi_A+\vpi_B)&=&\d(\vp+\vq)\d_{AB}\ ,\\
\d(\vp+\vq+\vpi_A+\vpi_B+\vpi_{\h'_j})&=&\d(\vp+\vq)X^j_{AB}\ ,\nonumber
\end{eqnarray}
where the second of these equations defines three 
symmetric matrices $X^j$.  In the Hamiltonian
these matrices occur in combination with the factors $e^{i\p_{Bj}}$ coming from $e^{i\ell_j}$,
and we define another set of matrices
\begin{equation}
\label{alphas}
\a^j_{AB}=X^j_{AB}\,e^{i\p_{Bj}}\ .
\end{equation}
The matrices $\a^j$ are real, and antisymmetric:
\begin{equation}
\label{antisymm}
\a^j_{BA}=X^j_{BA}\,e^{i\p_{Aj}}=X^j_{AB}\,e^{i(\p_B+\p_{\h'_j})_j}=-X^j_{AB}\,e^{i\p_{Bj}}
=-\a^j_{AB}\ ,
\end{equation}
because $(\p_{\h'_j})_j=\p$.  Hence the $\a^j$ are anti-Hermitian.  Using all this, we can simplify 
the expression for the Hamiltonian to
\begin{eqnarray}
\label{freeHfinal}
H&=&\int_p\Biggl\{\frac{1}{2}m\left(a_1^\dagger(\vp)a_1(\vp)+a_1(\vp)a_1^\dagger(\vp)
+a_2^\dagger(\vp)a_2(\vp)+a_2(\vp)a_2^\dagger(\vp)\right)\\
&&\hspace{0.7cm}+i\sum_j\sin{(p_j)}\left(a_1(\vp)\a^j a_2(\vp)-a_2^\dagger\a^j a_1^\dagger(\vp)\right)\Biggr\}\ .
\nonumber
\end{eqnarray}
Note that the integral over $\vp$ is over
the reduced Brillouin zone.

Finally, the eigenvalues of the $8\times 8$ matrix $\sum_j\sin{(p_j)}\a^j$ are equal to 
$\pm is(\vp)$ with $s^2(\vp)=\sum_j\sin^2{(p_j)}$.  Dropping a constant proportional
to $1/a^3$,
we can thus
write $H$ as a sum and integral over terms of the form
\begin{equation}
\label{hredef}
h(\vp)=m\left(a_1^\dagger(\vp)a_1(\vp)+a_2^\dagger(\vp)a_2(\vp)\right)
\pm s(\vp)\left(a_1(\vp)a_2(\vp)-a_2^\dagger(\vp)a_1^\dagger(\vp)\right)\ .
\end{equation}

\subsection{\label{eigen} Eigenvalues and eigenstates}
Dropping the dependence on $\vp$ in Eq.~(\ref{hredef}), our next step is
to find eigenvalues and left and right eigenstates of the non-Hermitian
Hamiltonian
\begin{equation}
\label{h1}
h=m(a_1^\dagger a_1+a_2^\dagger a_2)+s(a_1 a_2-a_2^\dagger a_1^\dagger)\ ,
\end{equation}
in which $a_{1,2}$ and $a^\dagger_{1,2}$ are a set of bosonic annihilation and creation
operators, $m>0$ and $s$ is real.  (Taking $s$ real without restriction on its sign takes care
of both signs in Eq.~(\ref{hredef}).)   Adapting the method of Ref.~\cite{Swanson}, we introduce new operators
\begin{eqnarray}
\label{newop}
b_1&=&\cos\theta\; a_1-\sin\theta\; a_2^\dagger\ ,\\
b_2&=&\cos\theta\;a_2-\sin\theta\; a_1^\dagger\ ,\nonumber\\
\tb_1&=&\cos\theta\;a_1^\dagger+\sin\theta\; a_2\ ,\nonumber\\
\tb_2&=&\cos\theta\;a_2^\dagger+\sin\theta\; a_1\ ,\nonumber
\end{eqnarray}
which obey the commutation rules
\begin{equation}
\label{comm}
[b_\a,\tb_\b]=\d_{\a\b}\ ,
\end{equation}
while all other commutators vanish.   Note that the operators $\tb_\a$
are not the Hermitian conjugates of the operators $b_\a$, which is
why this is a  ``generalized'' Bogoliubov transformation.
Expressed in terms of these operators $h$ becomes
\begin{eqnarray}
\label{hinb}
h&=&\left(m\cos(2\theta)+s\sin(2\theta)\right)(\tb_1 b_1+\tb_2 b_2)\\
&&\hspace{1.7cm}+\left(-m\sin(2\theta)+s\cos(2\theta)\right)(b_1 b_2-\tb_2\tb_1)+\mbox{constant}\ .\nonumber
\end{eqnarray}
Requiring the term proportional to $b_1 b_2-\tb_2\tb_1$ to vanish yields
\begin{equation}
\label{diag}
\theta =\frac{1}{2}\arctan(s/m)\ ,
\end{equation}
where we picked the solution that vanishes for $s\to 0$.  Substituting this solution into Eq.~(\ref{hinb})
gives
\begin{equation}
\label{hfinal}
h=\sqrt{m^2+s^2}\,(\tb_1 b_1+\tb_2 b_2)+\mbox{constant}\ .
\end{equation}
Even though $\tb_\a\ne b_\a^\dagger$, the operators $N_\a=\tb_\a b_\a$
are still number operators, and we can find a set of right eigenstates
$|n_1,n_2\rangle_R$ such that 
\begin{eqnarray}
\label{eigstates}
N_1|n_1,n_2\rangle_R&=&n_1|n_1,n_2\rangle_R\ ,\\
N_2|n_1,n_2\rangle_R&=&n_2|n_1,n_2\rangle_R\ ,\nonumber
\end{eqnarray}
and
\begin{eqnarray}
\label{lr}
b_1 |n_1,n_2\rangle_R\propto |n_1-1,n_2\rangle_R\ ,\\
b_2 |n_1,n_2\rangle_R\propto |n_1,n_2-1\rangle_R\ .\nonumber
\end{eqnarray}
Because of the upper bound~(\ref{Tbound}) on the absolute values of the
eigenvalues of $\hT_G$, which implies a lower bound on the real
part of the eigenvalues of $h$, we find that there exists a ``right vacuum''
state $|0,0\rangle_R$ annihilated by $b_1$ and $b_2$; otherwise 
$n_1$ and $n_2$ could be lowered indefinitely, violating this bound.
Therefore, it follows that $N_1$ and $N_2$ vanish on $|0,0\rangle_R$.   The right eigenstates
of $h$ are then given by
\begin{eqnarray}
\label{res}
|n,m\rangle_R&=&\frac{1}{\sqrt{n!m!}}\;\tb_1^n\tb_2^m|0,0\rangle_R\ ,\\
h|n,m\rangle_R&=&(n+m)E|n,m\rangle_R\ ,\qquad E\equiv\sqrt{m^2+s^2}\ ,\nonumber
\end{eqnarray}
where we have dropped the constant in Eq.~(\ref{hfinal}). The normalization of the states
we have chosen is convenient but
arbitrary, since the only relevant normalization condition relates right to left states
(Eq.~(\ref{orthon}) below). 
A similar reasoning leads
to a left ground state ${}_L\langle0,0|$, and a construction of the left eigenstates
\begin{eqnarray}
\label{les}
{}_L\langle n,m|&=&{}_L\langle 0,0| \frac{1}{\sqrt{n!m!}}\;b_1^n b_2^m\ ,\\
{}_L\langle n,m|h&=&{}_L\langle n, m|E(n+m)\ ,\nonumber
\end{eqnarray}
where the normalization here follows from Eq.~(\ref{res}) if we demand that
\begin{equation}
\label{orthon}
{}_L\langle n_1, m_1|n_2,m_2\rangle_R=\d_{n_1,n_2}\;\d_{m_1,m_2}\ .
\end{equation}
We note that
\begin{equation}
\label{nottrue}
{}_L\langle n,m|\ne\left(|n,m\rangle_R\right)^\dagger
\end{equation}
because $\tb_\a\ne b_\a^\dagger$; there is no simple relation between left- and right-eigenstates.  However, we do have a completeness relation: 
\begin{equation}
\label{eq:completeness}
\sum_{n,m}|n,m\rangle_R\ {}_L\langle n,m|=1\ .
\end{equation}
Completeness is not obvious, since our Hamiltonian is not Hermitian.   We refer to App.~\ref{completeness} for a proof.  

Under PT symmetry, $h(\vec p,s)\to h(-\vec p,-s)$, 
so it is actually the sum $h(\vec p,s) + h(-\vec p,-s)$ that is
PT symmetric. Since, from Eq.~(\ref{hfinal}),
$h(\vec p,s)$ and $h(-\vec p,-s)$ have identical eigenvalues, PT symmetry implies
that those eigenvalues must either be real or come in
complex-conjugate pairs. In fact all the eigenvalues are real, as seen in Eq.~(\ref{res}),
and all energy eigenstates of the sum may be chosen to be eigenstates of PT.
In the PT literature \cite{Bender} this situation is referred to as 
``unbroken PT symmetry.'' This is somewhat different from standard field-theory usage
of the terms broken and unbroken symmetry, which refer to the properties of the
ground state only.

\subsection{\label{two-point} Two-point correlators}
The free ghost partition function $Z$ is the trace of the (Euclidean)
evolution operator, which we may write, using Eq.~(\ref{eq:completeness}), as
\begin{equation}
\label{freeZ}
Z=\sum_{n,m}{}_L\langle n,m|e^{-Th}|n,m\rangle_R=\sum_{n,m}e^{-(n+m)ET}=\frac{1}{\left(1-e^{-ET}\right)^2}\ .
\end{equation}
{}From Eqs.~(\ref{res}),~(\ref{les}) and the commutation rules~(\ref{comm}),
it is straightforward to show that (with no sum over $\a$)
\begin{equation}
\label{btp1}
\langle b_\a(t)\tb_\a(0)\rangle=\frac{1}{Z}\sum_{n,m}{}_L\langle n,m|e^{-(T-t)h}
b_\a e^{-th}\tb_\a|n,m\rangle_R
=\frac{e^{-Et}}{1-e^{-ET}}\ ,
\end{equation}
and likewise
\begin{equation}
\label{btp2}
\langle\tb_\a(t)b_\a(0)\rangle
=\frac{e^{-E(T-t)}}{1-e^{-ET}}\ ,
\end{equation}
where we used 
\begin{equation}
\label{sum}
\sum_{n=0}^\infty n\;e^{-nET}=-\frac{\partial\ }{\partial(ET)}\sum_{n=0}^\infty e^{-nET}
=-\frac{\partial\ }{\partial(ET)}\frac{1}{1-e^{-ET}}=\frac{e^{-ET}}{\left(1-e^{-ET}\right)^2}\ .
\end{equation}
More interesting are the two-point correlation functions involving the
original creation and annihilation operators $a_\a$ and $a_\a^\dagger$.
For these we find from Eqs.~(\ref{btp1}) and~(\ref{btp2}), using the inverse
of Eq.~(\ref{newop}),
\begin{equation}
\label{aa}
\langle a_i(t)a_j(0)\rangle=\d_{i+j,3}\ \frac{s}{2E}\ \frac{e^{-Et}+e^{-E(T-t)}}{1-e^{-ET}}
=-\langle a_i^\dagger(t)a_j^\dagger(0)\rangle\ ,
\end{equation}
and
\begin{subequations}
\label{aad}
\begin{eqnarray}
\langle a_i(t)a_j^\dagger(0)\rangle&=&\d_{ij}\left(\frac{E+m}{2E}\ \frac{e^{-Et}}{1-e^{-ET}}
-\frac{E-m}{2E}\ \frac{e^{-E(T-t)}}{1-e^{-ET}}\right)\ ,\label{aada}\\
\langle a_i^\dagger(t)a_j(0)\rangle&=&\d_{ij}\left(-\frac{E-m}{2E}\ \frac{e^{-Et}}{1-e^{-ET}}
+\frac{E+m}{2E}\ \frac{e^{-E(T-t)}}{1-e^{-ET}}\right)\ .\label{aadb}
\end{eqnarray}
\end{subequations}
Equation~(\ref{aadb}) is a clear indication of the violation of unitarity in this theory:  In the limit $T\to\infty$, this correlator is negative (for $s\not=0$).  
In a normal theory, it would be a sum of decaying exponentials times positive
coefficients.

An alternative, path-integral derivation of the correlators Eqs.~(\ref{aa}) and~(\ref{aad}) is given
in App.~\ref{path}.  That approach also makes possible a direct comparison between 
our treatment of this nonunitary theory, and a treatment
of a similar Hamiltonian 
in the PT symmetry literature \cite{Jones}, where unitarity is restored through 
a redefinition of the theory.

\section{\label{effective} The effective theory}
We now return to the interacting theory defined by the transfer matrix
$T_{total}$ given in Eq.~(\ref{stagt}).  We want to 
argue that the 
there exists a corresponding chiral effective theory, and that it is given
by PQChPT. We will follow the discussion of Ref.~\cite{HLfound} as
closely as possible, so we first give a brief overview of the arguments 
there. Two key ingredients are the identification
of the light degrees of freedom (which we will collectively refer to 
as ``pions'') and clustering.  
For each of these ingredients, an extension of 
the Vafa--Witten theorem \cite{VW} turns out to be useful, as
discussed in Sec.~\ref{VWarg}. In the case
of clustering, we are then able to close the 
loophole that remains after constructing a bounded transfer matrix:  we can
argue that there is a gap (for strictly positive quark and ghost masses)  
between the ground state and the lowest excited
state, so that clustering is in fact exponential. The fact that an
extended Vafa--Witten theorem allows one to 
identify the light degrees of freedom has already been noted
in  Refs.~\cite{ShSh,BGS}.
In the case of unquenched
QCD, Lorentz invariance also 
plays an important role \cite{HLfound}.   Our setting is Euclidean, and we  instead have hypercubic invariance, which we assume to enlarge to $O(4)$ in the continuum limit, as usual.   We use this to argue
in Sec.~\ref{O4} that our pions satisfy the expected dispersion relation.

We then put all the ingredients together to write down the effective theory
in Sec.~\ref{PQChPTsection}.   A subtlety absent in the unquenched theory arises
because, in our theory with arbitrary fermionic and ghost quark masses,
pion masses can in general be complex.   In Sec.~\ref{CPTsection}, we use CPT
invariance, already introduced in Sec.~\ref{complete}, in order to show that
the phases of LECs in our theory are the same as in normal ChPT.   This
does not preclude the occurrence of complex masses in the effective theory,
but the effective theory can now be used to investigate this issue in more
detail, as we do in Sec.~\ref{non-degenerate}. 

As before, we will assume that all quark masses are always positive, both in the mass-degenerate and mass-nondegenerate cases considered below, as this is the setting for which the path integral~(\ref{ghostZ}) is well-defined, and the arguments of Ref.~\cite{VW} apply.   For simplicity, we will also assume that the continuum limit has been taken, so that we can ignore the
peculiarities of the staggered quark formalism with respect to species
doubling.   In particular, our arguments will apply to QCD with any number of 
continuum quarks.

\subsection{\label{Leutwyler} Recapitulation of Leutwyler's arguments}

Reference~\cite{HLfound} attempts to justify standard ChPT
as the effective theory for low-energy QCD. The argument is, roughly speaking,
divided into two parts.  In the first, which is largely qualitative, it
is argued that there is an effective chiral description of QCD in terms
of a Lagrangian that describes pions being exchanged between local vertices.
This is based on a few observations and assumptions:

\begin{itemize}
\item[$\bullet$]{} Pions are the lightest particles, so the low energy theory
is dominated by pion exchange.  In Minkowski space, this is the statement of
``pion pole dominance.''
\item[$\bullet$]{} Clustering, either 
power-law (when pions are massless) or
exponential (when pions are massive), 
implies that the interaction
among pions in a local region of space is independent of what is going on
far away. Thus the local interactions of pions may be described by
vertices that do not depend of the particular Green's function being 
considered.
\item[$\bullet$]{} The vertices are assumed to be expandable in a power
series in the momenta of the pions.  Note that one needs to {\it assume}\/ 
this in the case of massless pions, since Green's functions are themselves
not expandable in a power of series in momenta, due to
the infrared singularities that would result.  The assumption is 
that, once the singularities are accounted for by the 
exchange of massless pions, the vertices may
be expanded. However, this assumption seems to be unnecessary
in the massive case, because no infrared singularities would result from 
a momentum expansion.
\item[$\bullet$]{} Because more than one pion may be exchanged between any two given
vertices, pion loops must be included.  The fact that these loops can be expressed as the
usual four-dimensional momentum integrations of a quantum field theory is not explained
in detail.  However, the point seems to be that the underlying theory has a complete
set of states, and the sum over (on shell) states can be turned into four-dimensional
loop integrals over  off-shell states, by the standard arguments that relate old-fashioned
perturbation theory to the Feynman diagram approach. (See, for example, 
Ref.~\cite{Sterman:1994ce},
Sec.~9.5.)
\end{itemize}

The second, and much more lengthy,  part of the
discussion in Ref.~\cite{HLfound}  
is a demonstration 
that the chiral Lagrangian, with vector sources inserted
that transform like gauge fields,
can be chosen to have local chiral symmetry (up to
anomalies, which must be dealt with separately).
The main ingredient here is the chiral Ward--Takahashi
identities.
This part of the argument is crucial for showing that the chiral
Lagrangian has the standard form of ChPT.


\subsection{\label{VWarg} Pion spectrum and clustering: extension of the Vafa--Witten theorem}

An extension of the Vafa--Witten theorem serves two purposes here.  First of
all, it shows that, due to spontaneous breaking of chiral symmetry, 
pions exist in PQQCD, so that any effective
theory must be based on the exchange and interaction of pions.  Second, it shows that
pions are the {\it only}\/ light particles (absent particles that are light for accidental
reasons, 
which cannot be ruled out except by assumption).  This allows us to argue that the
fall off of correlation functions with distance, which follows from the existence of
a bounded transfer matrix, is actually exponential.  In other words, the theory obeys
exponential clustering.

PQQCD contains unquenched QCD.  Concretely, this means that if we 
consider correlation functions of operators made out of sea quarks (and gluons) only, these correlation functions coincide exactly with those of QCD
with only sea quarks, \ie, unquenched QCD.   Therefore, we know that PQQCD has excitations
which correspond to pions made only out of sea quarks, and that all 
correlation functions made out of sea-pion operators behave as they should
in a healthy quantum field theory.

First, consider PQQCD in which all quark masses are equal; \ie, 
$m_s=m_v=m_g$.\footnote{For the rest of this article we will use the subscripts
$s$, $v$, and $g$ to refer to sea, valence, and ghost quarks, respectively.}
In this case, we have a vector flavor
symmetry group%
\footnote{Until further notice, we use for simplicity the language of the 
``fake symmetries'' introduced in Ref.~\cite{BGPQ}, which do not take into
account all the subtleties coming from the ghost sector,
rather than the correct symmetries
introduced in Refs.~\cite{ShSh,GSS,DOTV}.
The subtleties do not affect the general argument.
In Sec.~\ref{CPTsection}, where the details
of the chiral Lagrangian matter, we use the 
nonperturbatively correct form.
See Ref.~\cite{MGLH} for a review of the issues arising from the ghost sector.}
$SU(N_s+N_v|N_v)$, which, if it is 
unbroken, relates the two-point (and other) correlation functions of all pions
in the theory to each other, and thus implies that there is a fully degenerate
multiplet of pions in the adjoint representation of $SU(N_s+N_v|N_v)$.
As already observed in Ref.~\cite{ShSh}, the vector-flavor group   $SU(N_s+N_v|N_v)$ is unbroken because of an extension of the
Vafa--Witten theorem \cite{VW}.   

Reference\ \cite{VW} contains two
proofs.
The first proof is based on a consideration of quark bilinears, and goes through without modification in the partially quenched case.   It implies immediately that
valence condensates are equal to sea condensates, because with
$m_v=m_s$ there is in fact no distinction between these
two types of fermions.   One ingredient that is needed is that the measure
in the path integral is positive, but this is not changed by the fact that
the valence part of the fermion determinant is missing.   It is also straightforward 
to prove that no flavor-symmetry breaking can take
place between the valence and ghost sectors, at the level of order
parameters made out of quark bilinears \cite{BGS}.

As Ref.~\cite{VW} points out, spontaneous breaking of
flavor symmetry could occur without  quark-bilinear order parameters.
  A more general proof that it does not considers the current-current 
correlation functions for the conserved flavor currents.   
If spontaneous
symmetry breaking took place, these correlation functions would 
couple to the corresponding massless Goldstone excitations, and would therefore
show a power-like fall-off.
  The idea is to show that such correlation functions instead satisfy an
exponential bound that is uniform in the gauge-field configuration.
This implies that no Goldstone mesons exist, and thus that flavor 
symmetry is unbroken.   
Since this argument is based on the Ward--Takahashi identities for flavor symmetry,
the framework also applies to the Euclidean partially quenched theory, because these
identities can also be worked out from the Euclidean path integral.

In fact, Ref.~\cite{VW} establishes a somewhat more complicated bound by 
considering a smeared, gauge-invariant quark propagator.   But the key
point for our discussion is that this bound is obtained for a fixed
gauge-field background, and independent of that gauge-field
background.   Since the only difference between the standard QCD
case and the PQQCD case is the relative weight of all gauge-field
backgrounds in the Euclidean path integral, the analysis of Ref.~\cite{VW} 
carries over to the partially quenched case.   

Therefore, we conclude that
the Vafa--Witten theorem applies to the PQQCD case with fully
degenerate quarks \cite{ShSh}.  PQQCD contains a complete $SU(N_s+N_v|N_v)$ multiplet of 
pseudo-Goldstone mesons in the mass-degenerate
case.   As in full QCD, if there are no other excitations that are 
accidentally light, one can consider the low-energy regime, in which
this partially quenched pion multiplet contains the only light excitations below a certain
scale.  
Since we know that  the pions in the sea sector have a nonzero mass, all
pions in the partially quenched theory are massive, and thus all correlation functions
of pion operators fall off exponentially, with a rate equal to the
pion mass. 

In the partially quenched theory, there are other states, made for example from multiple valence quarks,
that have no analogue in the sea sector, so these states are not constrained by the Vafa--Witten argument above.  
The absence of spontaneous symmetry breaking means that there is no fundamental reason
for these states to be light, but we cannot eliminate the possibility that they are 
accidentally light or massless.
As in Ref.~\cite{HLfound}, this possibility can be excluded only
by assumption: we assume that the pions are the only light states in the
theory. 

With this additional assumption, 
all correlation functions must decay exponentially, which we may
call ``cluster-like.'' This result is supported by overwhelming
numerical evidence from lattice QCD computations. 
For true clustering it is necessary in addition that 
a vacuum state exists and that this state is nondegenerate.   The
existence of a complete set of states follows from the existence of a 
transfer matrix for the partially quenched theory, \seef\ Sec.~\ref{complete}.\footnote{At least 
this is true at nonzero lattice spacing; we will ignore subtleties with defining a Hilbert space in the
continuum limit.  We also will follow Sec.~\ref{properties} in assuming that any
gauge configuration on which $\hT_G$ does not have a complete set of
eigenstates can be ignored.}   A state with a maximal absolute eigenvalue of the 
transfer matrix must also exist, because the transfer matrix is bounded.   We cannot,
however, prove uniqueness.   For example, we must by assumption
exclude the
possibility
that some breaking of a discrete symmetry (\eg\ parity) occurs, resulting
in two vacua, but without Goldstone bosons.  With these
assumptions, which are closely analogous to the assumptions required in Ref.~\cite{HLfound},  it follows that the degenerate partially quenched theory obeys exponential clustering.

Next, we consider the nondegenerate case, always keeping all quark
masses nonzero.    When we move away from the degenerate point by taking the valence, ghost,
and sea masses unequal (including, but not limited to, the partially quenched case where valence
and ghost masses remain degenerate),  we can still apply the bounds of
Ref.~\cite{VW} on
correlation functions 
directly, even if the vector symmetries are broken explicitly by the mass differences.  The gauge measure remains positive
in this case because it is just a product of two positive determinants (valence and
sea determinants) divided by the positive ghost determinant.  Thus the bounds go through,
and all connected correlation functions from point $x$ to point $y$ decay exponentially.
Here ``connected'' means they are formed out of one or more quark propagators that go
from $x$ to $y$.  For connected correlation functions, we thus automatically
have behavior which is ``cluster-like,'' in that correlators decay exponentially.

There are however many disconnected correlators in the partially quenched theory
(made using many valence flavors so that $x$ to $y$ contractions
do not occur),  
and once
again
most of these have no 
pure-sea analogue. 
We need to assume, as before, that such correlators do not
have power-law or anomalously light decay.  
With this assumption, the nondegenerate partially quenched theory has exponential
decay in all channels. If we further assume that there is a unique lowest state,
then the partially quenched theory obeys exponential clustering, 
even with nondegenerate masses.   

\subsection{\label{O4} The dispersion relation}
We will assume that the continuum limit of the 
Euclidean partially quenched theory has $O(4)$ ``space-time'' invariance.
What we wish to argue next is that this leads to the expected form of
propagators for one-particle pion states, and thus to the usual relation
between energy, mass and spatial momentum.  

The transfer matrix for the full theory has (right) eigenstates, which can
be classified according to their eigenvalues.   In addition, since the
theory is invariant under spatial translations, spatial momentum
is conserved, and the eigenstates of the transfer matrix can simultaneously
be labeled by their spatial momentum.   This follows because
the transfer matrix, even if it is
not Hermitian, generates translations in the time direction, and thus commutes
with the generators of spatial translations, \ie, spatial momentum.

Consider first a pion two-point correlator with zero spatial momentum. 
  We know from the preceding section that this correlator falls 
off  exponentially for large times:
\begin{equation}
\label{picorr}
C_\p(t)\propto e^{-m_\p|t|}\ .
\end{equation}
Here the parameter $m_\p$ might in principle be complex, since the Hamiltonian is
not Hermitian, and it is possible at this stage
that $m_\pi$ is one of a complex pair of eigenvalues.  
However we do know that  $m_\p$ has a positive real
part, as required by the  exponential damping of correlation functions.
There may also be other states that contribute to $C_\p(t)$, which would lead to
other exponentially damped contributions, with a faster decay rate (for
instance, three-pion states).   

The Fourier transform of $C_\p(t)$ is 
\begin{equation}
\label{FT}
f(p_4)=\int_{-\infty}^\infty dt\, e^{ip_4t-m_\p|t|}=\frac{2m_\p}{p_4^2+m_\p^2}\ .
\end{equation}
We now consider the correlator of a pion with a nonzero spatial momentum
$\vp$. By $O(4)$ invariance, the Fourier transform of this correlator has to be
equal to 
\begin{equation}
\label{f}
f(p)=\frac{2m_\p}{p^2+m_\p^2}\ ,
\end{equation}
where $p^2=\sum_\m p_\m p_\m$.  If we now Fourier transform back, 
we obtain the leading exponential of the pion correlator with nonzero
spatial momentum,
\begin{equation}
\label{picorrp}
C_\p(t,\vp)=\frac{1}{2\p}\int_{-\infty}^\infty dp_4\,\frac{2m_\p e^{-ip_4t}}{p_4^2+\vp^2+m_\p^2}=\frac{m_\p}{E}\,e^{-E|t|}\ ,
\end{equation}
with $E=\sqrt{m_\p^2+\vp^2}$.  It follows that the dispersion relation for pions in PQQCD
is the usual 
one, even though this theory is only defined in
Euclidean space and even though the ``mass'' $m_\p$ may in
principle be complex.

\subsection{\label{PQChPTsection} The chiral effective theory}
We now have all the needed ingredients to extend the arguments of Ref.~\cite{HLfound} to the partially quenched case.
The existence of the transfer matrix, together with the arguments and assumptions discussed
in Sec.~\ref{VWarg}, tell us that the theory clusters, and
that the lowest states are pions.  From Sec.~\ref{O4}, the pion states have arbitrary 
momentum, with the usual dispersion relation for energy in terms of spatial momentum
and (possibly complex) mass. Further, from Sec.~\ref{transfer}, the transfer matrix
has a complete set of eigenstates, under the mild assumption discussed
at the end of Sec.~\ref{complete}.

We then just follow Ref.~\cite{HLfound} step by step.  The low-energy theory is dominated
by exchange of the lightest particles, the pions  
({\it i.e.}, one has ``pion pole dominance,'' although of course
there are no poles in Euclidean space).  Pion interactions in a small
region of space are described by vertices, which do not depend on
the overall process due to clustering.  These vertices
are strictly local, {\it i.e.}, they
may be expanded in powers of momenta.  In fact, since we are not concerned here with
the massless case, this strict locality would appear to follow from the absence of
infrared singularities in Green's functions, and thus not require a separate assumption
as it does in the massless case considered in Ref.~\cite{HLfound}.
However, the relevant mass scale in this argument is the pion mass, $m_\p$,
and this argument would not exclude an expansion parameter 
$p/m_\pi$, with $p$ a typical momentum.   What we need instead is that vertices
can be expanded in $p/\Lambda_{QCD}$, and we thus end up having to make
the same assumption as in Ref.~\cite{HLfound}.
Long-distance correlation function then involve
the exchange of pions between local vertices.
Since the vertices are strictly local building blocks in correlation functions
in the effective theory,
two (or more) vertices can be joined by more than one 
pion propagator, leading to loops.  
And because the set of eigenstates is complete,
the correlation functions can be written as sums over intermediate states, 
which can
be turned into four-dimensional loop integrals as in the normal QCD case.
Note that the existence of double poles is not a problem for this argument,
because we can work with ghost and valence masses unequal, where we have only
single poles, and take the limit of equal ghost and valence masses only after the
loop expansion is in place. However, a subtlety could arise if the possibility of complex
masses is realized, because the standard arguments to relate sums over 
intermediate (on-shell) states to four-dimensional loop integrals would seem
to depend on knowing the locations of poles and cuts in Green's functions, and these
singularities will not be in the normal places if there are complex masses. We will assume
that no complex masses arise, and then note that this assumption
is confirmed {\it a posteriori}\/ by the effective theory of PQChPT 
both in the conventional (and most important) limit of degenerate valence-quark and 
ghost masses, and for appropriate choices of masses in the more general
theory that allow us to take that limit.  As discussed
in Sec.~\ref{non-degenerate}, however,  complex masses do seem to be possible
in PQChPT with some choices of nondegenerate valence- and ghost-quark masses.  Thus the reader
should keep in mind that the foundations of the effective theory are less
secure at present if such nondegenerate masses are allowed.

Following the first part of the discussion in  Ref.~\cite{HLfound}, 
the above ingredients (which we may roughly summarize as ``pions, vertices, and
loops''), are all that are needed 
to argue that there is 
an effective chiral theory describing the low energy behavior
of the theory. To get the {\it standard}\/ chiral theory (ordinary \ChPT\ in the full QCD case,
\PQChPT\ in the partially quenched case), a further technical argument, based
on the chiral  Ward--Takahashi identities, is needed 
to show that one may
choose the chiral theory to have {\it local} chiral symmetry. However, we claim
that this part of the argument goes through in the partially quenched case exactly as in the
full QCD case, since 
the partially quenched theory Ward--Takahashi identities are just like those
of the ordinary theory, but with the chiral group extended to a graded chiral group. 
The role of Lorentz invariance in this argument in Ref.~\cite{HLfound} is of course played
by Euclidean invariance in our case.

Note that we do not have to build the existence of double
poles into the chiral theory from the beginning, even though they can be shown to
occur already at the fundamental PQQCD level when valence and ghost masses are equal \cite{ShSh} .   Their existence in the appropriate limit  follows automatically from the effective theory, PQChPT.

\subsection{\label{CPTsection} The effective theory and constraints from CPT}

We now work with the nonperturbatively-correct partially quenched
chiral Lagrangian introduced
in Refs.~\cite{ShSh,GSS,DOTV}. 
The Lagrangian 
is a function of the chiral
fields $\Sigma(x)$ and $\Sigma^{-1}(x)$. $\Sigma$ is parameterized as
\begin{equation}
\label{Sigma}
\Sigma(x) = \exp( 2\Phi(x)/f)\;; \qquad \Phi(x) = \left(\begin{array}{cc}
i\phi(x)  & \omega(x) \\*
\bar\omega(x) & \tilde\phi(x) \end{array}\right) \,
\end{equation}
where $f$ is the pion decay constant, and where $\phi(x)$ are the quark-antiquark mesons, $\tilde\phi$ are the
ghost-antighost mesons, and $\omega$ and $\bar\omega$ are quark-antighost
and ghost-antiquark mesons, respectively.  While $\phi$ and $\tilde\phi$
are commuting fields, $\omega$ and $\bar\omega$ are Grassmann valued.
At leading order (and in the continuum limit), 
the Euclidean effective Lagrangian is \cite{GSS}
\begin{equation}
\cl_{\rm eff}=\frac{f^2}{8} \;
\str(\partial_\mu\Sigma\partial_\mu\Sigma^{-1})
-v \;\str(M\Sigma+\Sigma^{-1} M^\dagger) \label{Leff}
\end{equation}
where $\str$ denotes the supertrace, $M$ is the quark and ghost mass matrix,
and $v$ is a LEC.
We have assumed that the super-$\eta'$ field $\Phi_0\equiv -i\, \str\ln\Sigma = \tr(\phi +i\tilde\phi)$ has been integrated out, which is possible as long
as the theory is not completely quenched \cite{ShSh}.

In the past, it has been assumed that the LECs in the 
partially quenched chiral Lagrangian are real, just as the corresponding
ones are in the ordinary Hermitian chiral Lagrangian for full, unitary QCD.
For most LECs, such as $v$ above, this follows
from the fact that QCD is a special case of PQQCD, and LECs that the two effective theories share (for the same number of sea-quark flavors) are equal \cite{ShSh2000}.  
However there are additional LECs that
are unique to the partially quenched theory, and vanish for pure sea quantities
\cite{Sharpe-VdWater2003}.
We would like to be able to argue that those LECs are real too, in order to see
that complex meson masses will in general not occur.\footnote{They may
still occur in the flavor-diagonal sector, as we will see in the next section.}
The antiunitary
CPT symmetry of the theory, introduced in Sec.~\ref{complete}, can be used
to show this.

We must first determine the transformation properties of $\phi$,
$\tilde\phi$, 
 $\omega$ and $\bar\omega$ under CPT symmetry.  This is a bit subtle since
the discrete symmetries of staggered quarks (or ghosts) include
additional discrete taste transformations (see for instance the discussion
of spatial inversion symmetry in Ref.~\cite{GS}).  Since we want the transformation laws of pseudoscalar
mesons under continuum CPT
(without additional taste transformations), we must look on the 
lattice at taste-singlet mesons.  

From Ref.~\cite{Sharpe:1993ur},\footnote{The following argument can also be
given in the language of Ref.~\cite{stagmes}.} 
a staggered quark-antiquark bilinear with spin $\Gamma$ and taste $\Xi$
is 
\begin{equation}
\label{bilinears}
\frac{1}{64}\sum_{A,B}\bar q(x+A)\;q(x+B)\tr(\Omega^\dagger(A)\;\Gamma\;\Omega(B)\;\Gamma_\Xi^\dagger)
\end{equation}
where $A$ and $B$ hypercube vectors (with components 1 or 0 in each of the
four directions), $\Gamma_\Xi = \Xi^*$, $\Omega(x)\equiv\gamma_1^{x_1}\gamma_2^{x_2}\gamma_3^{x_3} \gamma_4^{x_4}$, and we have
omitted gauge links for
simplicity.
For the quark-antiquark meson field $\phi_{jk}$, with $j$ and $k$ the quark
and antiquark flavors, respectively, we have: 
\begin{eqnarray}
\phi_{jk}(y)& \dot{=} &\frac{i}{64} \sum_{A,B}\bar q_k(x+A)\;q_j(x+B)
\tr(\Omega^\dagger(A)\;\g_5\;\Omega(B)) \nonumber \\
& =& 
\frac{i}{16}\sum_{A,B}\left(\delta_{A+B,D}\;
(-1)^{B_1+B_3}\; \bar q_k(x+A)\;q_j(x+B)\right)
\label{phi-quarks}
\end{eqnarray}
where $\dot{=}$ should be read as ``has the same renormalized matrix elements as,'' 
$D=(1,1,1,1)$ is the diagonal of a hypercube, and
$y\equiv x+D/2$ is defined for convenience to be the middle of the hypercube.
The taste-singlet pseudoscalar meson is thus created by a ``four-link''
operator that joins opposite corners of a hypercube.
The overall factor of $i$ in Eq.~(\ref{phi-quarks}) is crucial; it is required to
make the propagator $\langle \phi_{jk}(y)\; \phi_{kj}(y')\rangle$ positive, which
follows at the chiral level from the definitions Eqs.~(\ref{Sigma}) and 
~(\ref{Leff}).%
\footnote{The positivity of this propagator in the continuum limit can
be most easily seen by noting that the corresponding propagator for taste
$\xi_5$ (taste-pseudoscalar) pions is a sum of absolute squares, which in
turn follows
from the fact that the staggered Dirac operator obeys $\cd^\dagger =
\epsilon \cd \epsilon$, where $\epsilon$ is a diagonal matrix in 
position space with $\epsilon(x)=(-1)^{x_1+x_2+x_3+x_4}$
along the diagonal.
The two factors of $i$ in the propagator cancel the minus sign from
Fermi statistics. In the continuum limit, the propagators for (flavor-charged)
mesons of arbitrary taste will be equal.}

For ghost-ghost mesons, the corresponding relation is
\begin{equation}
\tilde\phi_{jk}(y)\ \ \dot{=}\ \ 
\frac{1}{16}\sum_{A,B}\left(\delta_{A+B,D}\;
(-1)^{B_1+B_3}\; \chi^\dagger_k(x+A)\;\chi_j(x+B)\right)\ .
\label{tildephi-ghosts}
\end{equation}
Here there is no factor of $i$ because there is no minus sign in the propagator
due to statistics; by the definitions in Eqs.~(\ref{Sigma}) and ~(\ref{Leff}),
$\phi$ and $\tilde\phi$ have equal propagators.  Similarly the quark-antiqhost
and ghost-antighost mesons are given by 
\begin{eqnarray}
\omega_{jk}(y)& \dot{=} &
\frac{1}{16}\sum_{A,B}\left(\delta_{A+B,D}\;
(-1)^{B_1+B_3}\; \chi^\dagger_k(x+A)\;q_j(x+B)\right) \nonumber\\
\bar\omega_{jk}(y)& \dot{=} &
\frac{1}{16}\sum_{A,B}\left(\delta_{A+B,D}\;
(-1)^{B_1+B_3}\; \bar q_k(x+A)\;\chi_j(x+B)\right) \ .
\label{quark-antighost}
\end{eqnarray}
Again, no factor of $i$ is needed here to make the
$\langle \omega_{jk}(y)\;\bar\omega_{kj}(y')\rangle$ propagator have the
same sign as the corresponding $\phi$ propagator, which is required by
Eqs.~(\ref{Sigma}) and ~(\ref{Leff}).

We can now determine the CPT transformation rules for the meson fields by 
using the ghost and quark transformation rules, 
Eqs.~(\ref{CPT-ghost}) and ~(\ref{CPT-quark}). We find
\begin{eqnarray}
{\rm CPT}:\quad \phi(\vec y,y_4) &\to& \phi^{\rm T}(-\vec y,y_4)\ , \nonumber\\
{\rm CPT}:\quad \tilde\phi(\vec y,y_4) &\to& -\,\tilde\phi^{\rm T}(-\vec y,y_4)\ , \nonumber\\
{\rm CPT}:\quad \omega(\vec y,y_4) &\to& -\,\bar\omega^{\rm T}(-\vec y,y_4)\ , \\
\label{CPT-mesons}
{\rm CPT}:\quad \bar\omega(\vec y,y_4) &\to& \omega^{\rm T}(-\vec y,y_4)\ , \nonumber
\end{eqnarray}
with T the transpose, which acts on flavor indices. The factors of $(-1)^{x_4}$
in Eqs.~(\ref{CPT-ghost}) and ~(\ref{CPT-quark}), combined with the fact that the
spin-1/2 fields in the taste-singlet mesons are on different time-slices,
are important to getting the above signs under CPT.
The fact that the quark-antiquark meson field is even under CPT is standard.

From the definition of $\Phi$ in Eq.~(\ref{Sigma}) we then have
\begin{equation}
{\rm CPT}:\quad \Phi(\vec x,x_4) \to -\Phi^{\rm t}(-\vec x,x_4) \,
\label{CPT-Phi}
\end{equation}
where t denotes a graded transpose defined on a matrix with commuting
diagonal blocks and anticommuting off-diagonal blocks by
\begin{equation}
\left(\begin{array}{cc}
a  & b \\*
c  & d \end{array}\right)^{\rm t} = 
\left(\begin{array}{cc}
a^{\rm T}  & c^{\rm T} \\*
-b^{\rm T}  & d^{\rm T} \end{array}\right)\ .
\label{t-def}
\end{equation}
The operation t has the useful property that
\begin{equation}
\str\left(E^{\rm t}\,F^{\rm t}\,G^{\rm t}\,\dots\, K^{\rm t}\right) =
\str\left(\left[K\,\dots\, G\, F\, E\right]^{\rm T}\right) 
\label{transpose-rule}  
\end{equation}
where $E,F,G,K$ are matrices of the form in Eq.~(\ref{t-def}).  The property
can be proved by writing both sides 
as index sums,
counting the number of interchanges of anticommuting numbers between
the left and right sides, and showing that the sign resulting
from the interchanges is the same as the sign coming from the factors of
$-1$ in the definition of t. Note that, unlike the normal rule for the
transpose of a product of matrices with commuting entries, this equality is only true under the
supertrace. It is not true for the matrices themselves, as is already
obvious from considering a single matrix instead of a product.

We can now investigate the consequences of CPT for the chiral theory. Our definitions
have ensured that antiunitary CPT symmetry leaves the Euclidean action of the fundamental
theory invariant, so it will also leave the Euclidean chiral action invariant.
We will see that this implies
that the phases of all LECs in the partially quenched chiral Lagrangian are the same as
they would be in normal ChPT.  For a term that is special to the partially
quenched case (because in the normal case with the given
number of sea-quark flavors  it is not independent due to Cayley-Hamilton relations), 
the phase of the LEC is the same as it would be in the
ChPT theory for a QCD-like theory with sufficient numbers of flavors to make the
term independent.
The argument goes as follows:
\begin{itemize}
\item[1.\ ]{} 
Since Euclidean rotational invariance requires derivative operators to be
contracted as $\partial_\mu\partial_\mu$ 
we may redefine $\vec x\to-\vec x$ on the right-hand-side of Eq.~(\ref{CPT-Phi}) without
changing sign of any term. 
(We postpone discussion of anomaly terms, which may have the form
$\epsilon_{\mu\nu\lambda\sigma}\partial_\mu\partial_\nu\partial_\lambda\partial_\sigma$,
until item 5, below.)

\item[2.\ ]{} Chiral symmetry demands that any term in the Lagrangian be formed 
from one or more supertraces.  Therefore Eq.~(\ref{CPT-Phi}) and the rule 
Eq.~(\ref{transpose-rule}) mean that CPT effectively interchanges $\Sigma$ and
$\Sigma^{-1}$ and inverts the order of products of arbitrary numbers
$\Sigma$ and $\Sigma^{-1}$ matrices. The overall transpose on the right-hand side
of Eq.~(\ref{transpose-rule}) of course has no effect inside a supertrace.

\item[3.\ ]{} Because CPT is antiunitary it complex-conjugates all LECs.

\item[4.\ ]{} Thus the effect of CPT on the Lagrangian is exactly the same
as the effect of taking the Hermitian conjugate would be in 
a theory where $\Sigma$ is unitary. The phases are therefore the same
as in normal ChPT.

\item[5.\ ]{} For anomaly terms with 
$\epsilon_{\mu\nu\lambda\sigma}\partial_\mu\partial_\nu\partial_\lambda\partial_\sigma$,
an extra minus sign results from taking $\vec x\to-\vec x$, since we are not flipping
the sign of $x_4$ in Euclidean space.  Therefore such terms require an overall factor of $i$
(relative to what they would have in Minkowski space) in order to be CPT invariant.  However,
this factor of $i$ is precisely the factor that automatically arises in going from
Minkowski space to Euclidean space in any term with a single time derivative. Thus the LEC
again has the same phase as it would be ordinary ChPT.
\end{itemize}

A consequence of this result is that all parameters in the chiral effective
Lagrangian corresponding to the square of a meson mass are real
(in the usual convention in which they would also be real in the case
of a normal theory). Further, all mass terms shared by the PQ and normal
theory must have signs that ensure that squared meson masses are positive. This means
that as one turns on partial quenching by moving away from a point where valence
and ghost masses are degenerate with sea quark masses, the squared meson mass terms must
stay positive, at least until the valence, ghost and sea mass differences become large.%
\footnote{Negative squared masses would not imply imaginary masses, but rather
signal a phase transition and the need to find a new vacuum state. We believe such transitions, driven by possible
mass terms special to the partially quenched theory, 
cannot occur in the usual PQ case of degenerate valence and ghost masses, since the
dynamics of such a theory are controlled by the sea masses. We have not however ruled out
the existence of such transitions
when valence and ghost mass differences are large.}
As we will see in the next section, this still does {\it not} 
preclude complex masses from appearing in the theory for small perturbations, due to 
subtleties in the flavor-diagonal sector.   However, ``trivial'' complex masses,
coming directly from complex LECs, are ruled out. 

\subsection{\label{non-degenerate} Masses in the chiral theory}
As mentioned in Sec.~\ref{PQChPTsection}, the possibility of
complex masses is a concern for our derivation.  Once 
we have the chiral effective theory, however,
we can study this possibility in more detail.

Consider, for instance, the two-point function of a flavor-diagonal valence pion
and a flavor-diagonal ghost pion, in the mass nondegenerate case.
Extending the results of Ref.~\cite{BGPQ},
this two-point function is given by 
\begin{equation}
\label{vgprop}
D_{vg}(p)=-\frac{p^2+M_s^2}{N_v(p^2+M_s^2)(M_g^2-M_v^2)
+N_s(p^2+M_v^2)(p^2+M_g^2)}\ ,
\end{equation}
where we have chosen the number of ghost flavors, $N_g$,
equal to the number of valence flavors, $N_v$, while allowing $m_g\not=m_v$, and where we have sent the singlet part of the $\eta'$
mass to infinity.   $M_{v,g,s}$ are the masses of (flavor nondiagonal) valence,
ghost, and sea pions; in this subsection we use capital letters to distinguish
meson masses from quark masses.
The denominator of this expression
is a quadratic form in $-p^2$ of the form
\begin{equation}
\label{qform}
A(-p^2)^2+B(-p^2)+C\ ,
\end{equation}
with
\begin{eqnarray}
\label{ABC}
A&=&N_s\ ,\\
B&=&-N_v(M_g^2-M_v^2)-N_s(M_g^2+M_v^2)\ ,\nonumber\\
C&=&N_vM_s^2(M_g^2-M_v^2)+N_s M_g^2 M_v^2\ .\nonumber
\end{eqnarray}
The discriminant of this quadratic form is
\begin{equation}
\label{dis}
B^2-4AC=(N_v+N_s)^2(M_g^2-M_v^2)^2+4N_vN_s(M_g^2-M_v^2)(M_v^2-M_s^2)\ .
\end{equation}
Note that the discriminant vanishes in the limit
$M_g=M_v$, and Eq.~(\ref{vgprop}) has a double pole
in that case as expected, unless also $M_s=M_v$.  The double pole is coming
from cancellations between ghost and valence terms, which arise in turn because
``wrong sign'' contributions from the ghosts due ultimately to their non-Hermitian Hamiltonian.

In order to obtain only real single poles, 
$B^2-4AC$ needs to be positive, and this turns out not to be always the case.
For simplicity, take $N_v=N_s\equiv N$, so that
\begin{equation}
\label{equalN}
\left(B^2-4AC\right)|_{N_v=N_s=N}=4N^2(M_g^2-M_v^2)(M_g^2-M_s^2)\ .
\end{equation}
This is only positive if $M_g<M_v$ and $M_g<M_s$, or if
$M_g>M_v$ and $M_g>M_s$.   This means that even a small
perturbation from the degenerate case can lead to a situation in
which the discriminant is negative, leading to a conjugate pair of
complex zeros of Eq.~(\ref{qform}).   The real part of these
zeros is given by $-B/(2A)=M_g^2$.   
The two-point function is
still well defined, but the effective theory implies that it is possible to be in the situation
that energies of states become complex (with a positive real part).
Note that the existence of a conjugate pair of energies
is consistent with the PT symmetry of the theory. The reality of $M^2_g$,
$M^2_v$ and $M^2_s$, which 
follows from the reality of the LECs
of theory, does however not guarantee that the  poles
of this real propagator
occur at real values of $-p^2$.  We emphasize, though, that it is
always possible to take the limit $M_g\to M_v$ in such a way that the energies remain real.
In particular, if $M_v < M_s$, we can take  $M_g\to M_v$ from below, and if  $M_v > M_s$, we
can take $M_g\to M_v$ from above.

Another instructive example is to take $M_v=M_s\ne M_g$ 
(while leaving
$N_v$ and $N_s$ arbitrary), which keeps
$B^2-4AC>0$.   Now the zeros $-p^2_\pm$ of the quadratic form are
\begin{eqnarray}
\label{zeroes}
-p^2_-&=&M_v^2\ ,\\
-p^2_+&=&M_g^2+\frac{N_v}{N_s}(M_g^2-M_v^2)\ .\nonumber
\end{eqnarray}
However, the zero $-p_+^2$ becomes negative for 
\begin{equation}
\label{cond}
M_v^2>M_g^2\left(1+\frac{N_s}{N_v}\right)>M_g^2\ .
\end{equation}
This would make the Fourier transform of Eq.~(\ref{vgprop}) ill-defined,
but this can only happen if one perturbs $M_v$ far enough away from
$M_g$.   This suggests that for a choice of $M_v$ such that 
$M_v^2=M_g^2(1+N_s/N_v)$ a phase transition takes place. 
In that case, 
the effective partition function would have to be evaluated by performing
a different saddle-point expansion than the one assumed in writing down
Eq.~(\ref{vgprop}).
Of course, as already remarked above, no such problems,
and no complex energies, appear when $M_g=M_v$, which is the
``physical'' case of PQQCD, because in that case $B^2-4AC=0$,
and we recover the usual double pole.   We therefore do not pursue the possibilities arising for $M_g\ne M_v$ further. 

\section{\label{conclusion} Conclusion}
In this article, we have presented the theoretical evidence that \PQChPT\ provides the correct low-energy effective theory
for PQQCD.   Our starting point is the discussion by Leutwyler \cite{HLfound}
of the foundations of chiral perturbation theory in the case of full QCD.
The cluster property of a Lorentz invariant, local quantum 
field theory plays a central role in Ref.~\cite{HLfound}; unitarity, in contrast,
 appears not to be needed.   This starting point is essential in any attempt to extend the validity of \ChPT\ to 
the partially quenched case, in which unitarity is lost.   Therefore, our main
task has been to see to what extent the cluster property can be established in
PQQCD as well.   

The key ingredients are the existence of a bounded transfer matrix, as well
as an extension of the Vafa--Witten theorem to the partially quenched case.
The existence of the transfer matrix allows us to identify a complete set of states in the theory, even though the transfer matrix and 
the corresponding states do not have all the usual
properties they possess in a unitary theory.  An important point,
however, is that the pion states do satisfy a rotationally invariant dispersion
relation.
The Vafa--Witten theorem \cite{VW}
allows us to connect the space-time dependence of correlation functions 
in the partially quenched theory with those of correlation functions of
the corresponding full QCD theory with the same set of sea quarks.
While these ingredients together are not sufficient to provide a rigorous
proof of clustering, we have identified the further assumptions needed to establish the cluster property in PQQCD, assuming that it holds in the full theory.\footnote{We are not aware of a rigorous proof of the cluster property for
full QCD either; for a discussion of the cluster property in quantum field theories, see Ref.~\cite{SWbook}.}

Once the cluster property is established, the argument for the correctness of \PQChPT\ as 
the low-energy effective theory for PQQCD follows mainly along the same
lines as that given in Ref.~\cite{HLfound}.   While in both cases some additional
assumptions are needed (such as the assumption that no other
accidentally light states exist in the theory), these additional assumptions in general
have little to do with the ``sickness'' of the partially quenched theory,
and we thus believe them to be equally plausible in both the full and
partially quenched cases. 
One issue unique to the partially quenched case is the
possibility of complex energies, which could make it difficult to write effective-theory 
loops as normal four-dimensional integrals, and require instead three-dimensional
integrals over on-shell states.   The arguments presented in Sec.~\ref{effective}
support the assumption that this situation does
not occur in the only case of practical interest, in which ghost quark masses are chosen equal to the
corresponding valence quark masses, \ie, the case usually
referred to as partial quenching.   Section \ref{CPTsection} shows that CPT symmetry requires the effective theory to have real low energy constants.   
This restricts the problem of complex masses
to the flavor-diagonal sector.   As demonstrated in Sec.~\ref{non-degenerate}, 
complex masses (or poles located at negative values of (Euclidean) $-p^2$)
do not occur when all ghost quark masses are chosen equal to the
corresponding valence quark masses.   The only ``sickness'' is the familiar
occurrence of double poles, at positive values of $-p^2$, in the valence sector.

Our framework for the construction of the effective low-energy theory 
generalizes to the
fully nondegenerate case, in which ghost-quark masses are not
equal to valence-quark masses.   It turns out that this generalized theory
contains some new properties absent in the partially quenched case; these
properties can be
investigated in 
the corresponding effective field theory.
In particular, one can choose values for ghost and valence masses
such that complex poles appear in the disconnected propagator.
We leave open the question of the validity of the standard loop
expansion of the effective theory for this case.
Of course, as already emphasized above,
this issue is primarily of 
academic interest, since, by construction, ghost and valence quark
masses are always equal in numerical applications of PQQCD.

\vspace{3ex}
\noindent {\bf Acknowledgments}
\vspace{3ex}

We thank Carl Bender and Michael Ogilvie for discussions on PT symmetry,
and Steve Sharpe for reading 
the manuscript and for offering many helpful
suggestions for improving it.
We are
grateful to the Galileo Galilei Institute for Theoretical Physics and
the INFN, whose
hospitality and support
allowed us to complete this work. This work was also supported in part by the US Department of Energy under grants
DE-FG02-91ER40628 (CB) and DE-FG03-92ER40711 (MG).  In
addition, MG is supported in part
by the Spanish Ministerio de Educaci\'on, Cultura y Deporte, under program SAB2011-0074.

\appendix
\section{\label{completeness} Completeness}
In this appendix, we prove completeness in the free theory
of the basis of right eigenstates given by Eq.~(\ref{res}),
or the left eigenstates given by Eq.~(\ref{les}).
The argument has three steps.   First, we prove that  $|0,0\rangle_R$,
${}_L\langle{0,0}|$
are the unique right and left vacuum states.  We then show that 
all right eigenstates are obtained by acting with the raising operators $\tb_1$ and $\tb_2$
on $|0,0\rangle_R$ (and similarly for left eigenstates). Finally, we prove that that no blocks of the form
of the form~(\ref{Jordan}), or larger generalizations thereof, occur in the
Jordan normal form of the Hamiltonian matrix of the free theory.

The uniqueness of the right and left vacua can be proved by working in position space
and solving the differential equations corresponding to the definitions of 
$|0,0\rangle_R$ and ${}_L\langle 0,0|$, which are given in operator form by
\begin{eqnarray}
\label{rvacuum}
b_1\,|0,0\rangle_R&=&(\cos\theta\ a_1-\sin\theta\ a_2^\dagger)\,|0,0\rangle_R=0\ ,\\
b_2\,|0,0\rangle_R&=&(\cos\theta\ a_2-\sin\theta\ a_1^\dagger)\,|0,0\rangle_R=0\ ,\nonumber
\end{eqnarray}
and
\begin{eqnarray}
\label{lvacuum}
{}_L\langle 0,0|\,\tb_1&=&{}_L\langle 0,0|\,(\cos\theta\ a_1^\dagger+\sin\theta\ a_2)=0\ ,\\
{}_L\langle 0,0|\,\tb_2&=&{}_L\langle 0,0|\,(\cos\theta\ a_2^\dagger+\sin\theta\ a_1)=0\ ,\nonumber
\end{eqnarray}
where we used Eq.~(\ref{newop}).

Using $x$ for the position associated with oscillator described by $a_1$, $a_1^\dagger$
and $y$ for the oscillator described by $a_2$, $a_2^\dagger$,
we have
\begin{eqnarray}
\label{a-x}
a_1 = \frac{1}{\sqrt{2}}(x + \frac{\partial }{\partial x}),&\qquad &
a^\dagger_1 = \frac{1}{\sqrt{2}}(x - \frac{\partial }{\partial x}) \\ 
a_2 = \frac{1}{\sqrt{2}}(y + \frac{\partial }{\partial y}),&\qquad &
a^\dagger_2 = \frac{1}{\sqrt{2}}(y - \frac{\partial }{\partial y})\ . \nonumber
\end{eqnarray}
Now let
\begin{equation}
\label{psi-chi-def}
\langle x,y|0,0\rangle_R = \psi(x,y), \qquad {}_L\langle 0, 0 |x,y\rangle_R = \chi^*(x,y)
\end{equation}
be the position-space wavefunctions for the vacua.
The equations for $\psi(x,y)$ and $\chi(x,y)$ are
\begin{eqnarray}
\label{psi-chi}
\left[\cos\theta\left(x+\frac{\partial}{\partial x}\right)
-\sin\theta\left(y-\frac{\partial}{\partial y}\right)\right]\psi(x,y)&=&0\ , \\
\left[-\sin\theta\left(x-\frac{\partial}{\partial x}\right)
+\cos\theta\left(y+\frac{\partial}{\partial y}\right)\right]\psi(x,y)&=&0\ ,\nonumber \\
\left[\cos\theta\left(x+\frac{\partial}{\partial x}\right)
+\sin\theta\left(y-\frac{\partial}{\partial y}\right)\right]\chi(x,y)&=&0\ ,\nonumber\\
\left[\sin\theta\left(x-\frac{\partial}{\partial x}\right)
+\cos\theta\left(y+\frac{\partial}{\partial y}\right)\right]\chi(x,y)&=&0\ .\nonumber
\end{eqnarray}
The solutions are (recalling Eq.~(\ref{diag}))
\begin{eqnarray}
\label{psi-chi-soln}
\psi(x,y)&=&A\,\mbox{exp}\left(-\frac{E}{2m}(x^2+y^2)+\frac{s}{m}\,xy\right)
\ , \nonumber \\
\chi(x,y)&=&A\,\mbox{exp}\left(-\frac{E}{2m}(x^2+y^2)-\frac{s}{m}\,xy\right)\ .
\end{eqnarray}
Since the only free parameter in these solutions is the normalization $A$, 
we have shown that the right and left vacua are unique.   

It is now clear (following the argument after Eq.~(\ref{lr}))
that any right eigenstate of $h$ can be repeatedly lowered
by application of $b_1$ and $b_2$ until 
we reach $|0,0\rangle_R$.  This is all we need in order to show
that the eigenstate can, in turn,  be obtained by operating with $\tb_1$ 
and $\tb_2$ on the vacuum, and is therefore just proportional to
one of the states $\rket{n,m}$ in Eq.~(\ref{res}).   For example, consider an eigenstate 
$|1',0'\rangle_R$, with
\begin{eqnarray}
\label{sexample}
\tb_1 b_1\,|1',0'\rangle_R&=&\,|1',0'\rangle_R\ ,\\
\tb_2 b_2\,|1',0'\rangle_R&=&0\ .\nonumber
\end{eqnarray}
By uniqueness of the vacuum,
\begin{equation}
\label{uu}
b_1\,|1',0'\rangle_R\propto |0,0\rangle_R\ .
\end{equation}
Operating on both sides of this equation with $\tb_1$ and using Eq.~(\ref{sexample}) then proves that
\begin{equation}
\label{uv}
|1',0'\rangle_R\propto\tb_1\,|0,0\rangle_R=|1,0\rangle_R\ .
\end{equation}
It is straightforward to extend this step to a proof by induction that,
similarly, 
\begin{equation}
\label{uvn}
|n',m'\rangle_R\propto|n,m\rangle_R\ ,
\end{equation}
for an eigenstate $|n',m'\rangle_R$ of the number operators $\tb_1 b_1$ and $\tb_2 b_2$ with eigenvalues $n'$ and $m'$ respectively.   An analogous
argument extends this first step to left eigenstates as well. 

The fact that we have shown that all right and left eigenstates are of the form 
of Eq.~(\ref{res}) and~(\ref{les}) does not yet quite prove completeness
of these states.   Since the Hamiltonian~(\ref{hfinal})
is not Hermitian, it is possible that its Jordan normal form has nontrivial
blocks of the form~(\ref{Jordan}), with $\k\ne 0$.   We now prove that this
does not happen in the free theory.   First, the Hamiltonian is proportional
to the sum of two number operators, $N_1=\tb_1 b_1$ and $N_2=\tb_2 b_2$, which commute with each other.   Therefore, the matrix representation
of the sum is a direct product of the two matrices representing $N_1$ and
$N_2$.   It is thus sufficient to show that no nontrivial Jordan blocks
occur in the matrix representation of $N=\tb b$, where $b=b_i$ and $\tb=\tb_i$, with $i=1$ or 
$2$.

Now assume that a hypothetical Jordan block of arbitrary size $p\times p$
occurs, with $p>1$.   Such blocks have an eigenvalue on the diagonal,
and an arbitrary constant $\k$ on the ``superdiagonal.''   The block is trivial if $\k=0$; nontrivial otherwise.
We already know that possible eigenvalues
of the number operator $N$
are $n=0,1,2,\dots$.  So, for example, a block for eigenvalue $n$ and
$p=4$ would look like
\begin{equation}
\label{Jordanp}
\left( \begin{array}{cccc}
n&\k& 0 & 0 \\
          0&n &\k& 0  \\
          0& 0 &n & \k  \\
	  0 &0 & 0 & n\end{array} \right) \ .
\end{equation}
We will prove by induction on $n$ that
no such blocks can exist for any $n$, and any size $p$.  For any such hypothetical
block, let $\rket{n}$ be the true right eigenvector and $\rket{A^{(n)}}$ be the first
nontrivial generalized right eigenvector.
These states obey
\begin{eqnarray}
\label{n-and-A}
N\,\rket{n} &=& n\rket{n} \ , \\
N\,\rket{A^{(n)}} &=& n\rket{A^{(n)}}+\k\rket{n} \ .\nonumber
\end{eqnarray}
For example, in the $p=4$ case above, $\rket{n}$ and $\rket{A^{(n)}}$ could
be represented by
\begin{equation}
\label{n-and-A-def}
\left(\begin{array}{c}1\cr 0 \cr 0 \cr 0\end{array} \right)\quad {\rm and}\quad 
\left(\begin{array}{c}\g\cr 1 \cr 0 \cr 0\end{array} \right) \ ,
\end{equation}
respectively, with $\g$ arbitrary.

The theorem that we will prove is:
\begin{center}
\begin{minipage}[h]{0.8\textwidth}
If, for any $n$, there exists a state  $\rket{A^{(n)}}$ that is 
linearly independent of the eigenvector $\rket{n}$ and obeys 

\vspace{-2mm}

\begin{equation}
\label{theorem}
N\,\rket{A^{(n)}} = n\rket{A^{(n)}}+\k\rket{n}\ ,
\end{equation}

\vspace{-2mm}
then $\kappa=0$.\\
\end{minipage}
\end{center}
Thus  $\rket{A^{(n)}}$ is a right eigenstate degenerate with $
\rket{n}$.  However, since we showed above that the eigenstates of $N$ are 
nondegenerate, $\rket{A^{(n)}}$ would have to be proportional to $
\rket{n}$, which contradicts the assumption of linear independence.
Thus, no state  $\rket{A^{(n)}}$ can exist, and the Jordan block is not only
trivial ($\k=0$) but is actually one-dimensional ($p=1$).

We first show the theorem is true for $n=0$. For a nontrivial Jordan block 
with $n=0$ we would
have
\begin{equation}
\label{theorem0a}
N\,\rket{A^{(0)}} = \k\rket{0}\ ,
\end{equation}
which implies
\begin{equation}
\label{theorem0b}
N b\,\rket{A^{(0)}} = (bN-b)\rket{A^{(0)}} = -b\,\rket{A^{(0)}}\ .
\end{equation}
So $b\,\rket{A^{(0)}}$ would have to be an eigenstate of $N$ with eigenvalue $-1$, which is impossible
since $N$ is positive semidefinite. Thus $b\,\rket{A^{(0)}}$ must vanish,
which implies
\begin{equation}
\label{theorem0c}
N \,\rket{A^{(0)}} = \tilde b b\,\rket{A^{(0)}} = 0 \ .
\end{equation}
Comparing with Eq.~(\ref{theorem0a}), we see that $\kappa=0$, as desired.

We now assume that the theorem is true for eigenvalue $n$, and prove it is also true for
eigenvalue $n+1$.
Suppose we have a state $\rket{A^{(n+1)}}$ that is linearly independent of $\rket{n+1}$
and satisfies
\begin{equation}
\label{theoremnp1a}
N\,\rket{A^{(n+1)}} = (n+1)\rket{A^{(n+1)}}+\k\rket{n+1}\ .
\end{equation}
Then
\begin{equation}
\label{theoremnp1b}
N b\,\rket{A^{(n+1)}} = (bN-b)\rket{A^{(n+1)}} = nb\,\rket{A^{(n+1)}}+\sqrt{n+1}\;\k\rket{n}\ .
\end{equation}
So the state $ b\rket{A^{(n+1)}}$ satisfies Eq.~(\ref{theorem}) for eigenvalue $n$, 
with $\k$ replaced by $\k'=\sqrt{n+1}\;\k$. Furthermore,  $ b\rket{A^{(n+1)}}$ is linearly independent of
$\rket{n}$.  If not,  $ b\,\rket{A^{(n+1)}}=\alpha\rket{n}$ for some constant $\alpha$,
which implies by Eq.~(\ref{theoremnp1a}) that
\begin{equation}
\label{theoremnp1c}
\tilde b b\,\rket{A^{(n+1)}} = (n+1)\rket{A^{(n+1)}}+\k\rket{n+1}= \tilde b\, \alpha\rket{n} = 
\sqrt{n+1}\;\alpha\rket{n+1}\ ,
\end{equation}
contradicting the assumption that  $\rket{A^{(n+1)}}$ is linearly independent of $\rket{n+1}$.  Therefore, $b\,\rket{A^{(n+1)}}$ satisfies the conditions of the induction hypothesis, which
implies $\k'=0$, and hence $\k=0$.

This concludes the proof of our theorem, and hence completeness of the
free theory as in Eq.~(\ref{eq:completeness}).

\section{\label{path} Path integral for the free ghost theory}
Starting from Eq.~(\ref{h1}), we reconstruct a path integral which yields the same
correlation functions as those generated by the transfer matrix defined in
terms of $h$. This is an alternative to the operator
analysis based on the generalized
Bogoliubov transformation, Eq.~(\ref{newop}).
For simplicity, we take $T\to\infty$ from the beginning.

We start by defining new
 fields $\psi_i$ and conjugate momenta $\rho_i$
\begin{equation}
\label{psi-rho}
\psi_i = \frac{1}{\sqrt{2m}}(a_i + a^\dagger_i)\ ,\qquad 
\rho_i = -i\sqrt{\frac{m}{2}}(a_i -a^\dagger_i)\ ,
\end{equation}
which are conventionally normalized,
\begin{equation}
\label{commutator}
[\psi_i, \rho_j] = i\,\delta_{ij} \ .
\end{equation}
The Hamiltonian then may be written
\begin{equation}
\label{h-psi-rho}
h = \frac{1}{2}\left(\rho_1^2+\rho_2^2 + m^2(\psi_1^2+\psi_2^2) \right)
+is\,(\psi_1\rho_2+\psi_2\rho_1) + {\rm constant} \ .
\end{equation}
Introducing sources $J_i$ for $\psi_i$ and $K_i$ for $\rho_i$, the partition function
in the Hamiltonian form of the path integral is
\begin{equation}
\label{PIham}
Z[J_i,K_i] = \int\cd\psi_1\; \cd\psi_2\; \cd\rho_1\; \cd\rho_2\; \exp(i\rho_j\dot\psi_j
-h + J_j \psi_j + K_j\rho_j) \ ,
\end{equation}
with a sum over the repeated index $j$, and with $\dot\psi_j\equiv d\psi_j/dt$.
The quadratic integrals over $\rho_1$ and $\rho_2$ are easily done,\footnote{The $\rho_j$ have open
boundary condition,
as is always the case in the path integral formulation.}
and the result is
\begin{equation}
\label{PIlag}
Z[J_i,K_i] = \int\cd\psi_1\; \cd\psi_2\; e^{-\cl_E[J_i,K_i]} \ ,
\end{equation}
with the Euclidean Lagrangian $\cl_E[J_i,K_i]$ given by
\begin{eqnarray}
\label{LE}
\cl_E[J_i,K_i] 
 &=& \frac{1}{2}\left(\dot\psi_1^2+\dot\psi_2^2 +  E^2(\psi_1^2+\psi_2^2) \right)
-iK_1(\dot\psi_i-s\psi_2) -iK_2(\dot\psi_2-s\psi_1) \nonumber\\
&& -J_1\psi_1  -J_2\psi_2 - \frac{1}{2}(K_1^2+K_2^2) -s\frac{d}{dt}(\psi_1\psi_2) \ .
\end{eqnarray}
The terms quadratic in $K_j$ are standard, and just give contact terms that we
drop from now on.  
We will also drop the final, total-derivative term, since we want to calculate a partition
function with periodic boundary conditions on $\psi_j$.
Note, however, that for
computing transition amplitudes between $\psi_j$ eigenstates, dropping the total derivative
would be incorrect; indeed that term leads to violations
of unitarity. Violations arise because the Minkowski-space
transition amplitude 
\begin{equation}
T_{ba}=\langle\psi_{1,b}\psi_{2,b}; t_b|\psi_{1,a}\psi_{2,a}; t_a\rangle
\end{equation} is
proportional to $\exp[s(\psi_{1,b}\psi_{2,b}-\psi_{1,a}\psi_{2,a})]$ and does therefore
does not obey $T_{ba}=T^*_{ab}$.

If we focused only on correlation functions of $\psi_j$, the
sources $K_j$ could also be dropped completely,
and one would immediately see that the correlation
functions are the completely normal correlators of two harmonic oscillators with frequency E.
One could in fact {\it change}\/ the rules at this point, and write down a standard (double)
harmonic oscillator Hamiltonian $h'$ that gives these same $\psi_j$ correlators:
\begin{equation}
\label{new-h}
h' = \frac{1}{2}\left(\rho_1^2+\rho_2^2 + E^2(\psi_1^2+\psi_2^2) \right)
\end{equation}
The analogous step for the Hamiltonian of Ref.~\cite{Swanson} is taken
in Ref.~\cite{Jones}. However, this is really a change of rules, because it changes the
correlators involving $\rho_j$. As we will see below, nonunitary effects 
we already obtained in Sec.~\ref{two-point} show up in 
those correlators.

From Eqs.~(\ref{PIlag}) and~(\ref{LE}), we can easily compute all two-point correlators by taking derivatives
with respect to $J_j$ and $K_j$. The results are: 
\begin{subequations}
\label{psipsi}
\begin{eqnarray}
\langle \psi_i(t)\psi_j(0)\rangle &=& \delta_{ij}\, \frac{e^{-Et}}{2E} 
\hspace{4.5truecm}  \left[= \delta_{ij}\,\frac{e^{-Et}}{2E}\right]\ ,
\label{psipsia}
\\
\langle \psi_i(t)\rho_j(0)\rangle &=& \delta_{ij}\,\frac{ie^{-Et}}{2} 
+ \delta_{i+j,3}\, \frac{-ise^{-Et}}{2E} 
\hspace{1.36truecm}  \left[= \delta_{ij}\,\frac{ie^{-Et}}{2}\right]\ ,
\label{psipsib}
\\
\langle \rho_i(t)\psi_j(0)\rangle &=& \delta_{ij}\,\frac{-ie^{-Et}}{2} 
+ \delta_{i+j,3}\, \frac{-ise^{-Et}}{2E} 
\hspace{1.03truecm}  \left[= \delta_{ij}\, \frac{-ie^{-Et}}{2}\right]\ ,
\label{psipsic}
\\
\langle \rho_i(t)\rho_j(0)\rangle &=& \delta_{ij}\, \frac{m^2e^{-Et}}{2E} 
\hspace{3.95truecm}  \left[= \delta_{ij}\, \frac{Ee^{-Et}}{2}\right]\ .
\label{psipsid}
\end{eqnarray}
\end{subequations}
Here, the corresponding results in the standard theory defined by $h'$ are shown in square
brackets.
For both our theory and the standard theory, 
these correlators satisfy the conditions required
by the commutators Eq.~(\ref{commutator}) in the limit $t\to0$. (The relevant correlators
are unaffected by the dropped contact terms.)  

Aside from the off-diagonal terms in
the $\psi$-$\rho$ correlators, the difference between our theory and the standard theory
appears in the normalization of the $\rho$-$\rho$ correlator in Eq.~(\ref{psipsid}).
 From Eq.~(\ref{psi-rho}) and Eq.~(\ref{psipsi}), 
we can easily check that we reproduce the correlators Eqs.~(\ref{aa}) and~(\ref{aad}) for
the creation and annihilation operators in the limit $T\to\infty$. In particular, the unitarity-violating
negative result  for $\langle a_i^\dagger(t) a_j(0)\rangle$ shows up here because of the
incomplete cancellation of the various contributions from Eqs.~(\ref{psipsia}) 
through~(\ref{psipsid}).  
(Note that, in order to recover expected results for correlators of
creation and annihilation operators in the standard theory, one must
replace the definitions Eq.~(\ref{psi-rho}) by letting $m\to E$, so that 
the Hamiltonian $h'$ takes its standard form in terms of $a_i$ and $a_i^\dagger$.)


\end{document}